\documentclass{aa}  
\usepackage{graphicx}
\usepackage{txfonts}
\begin{document}
   \title{VLT spectropolarimetry of the optical transient in 
    NGC~300\thanks{Based on observations made with ESO Telescopes at 
    Paranal Observatory under Program ID 281.D-5016.}}
  \subtitle{Evidence of asymmetry in the circumstellar dust}
   \author{F. Patat\inst{1}
   \and    J.R. Maund\inst{2}
   \and
    S. Benetti\inst{3}
    \and 
    M.T. Botticella\inst{4}
   \and
    E. Cappellaro\inst{3}
   \and
    A. Harutyunyan\inst{3,5}
   \and
    M. Turatto\inst{6}
}

   \offprints{F. Patat}

   \institute{European Organization for Astronomical Research in the 
	Southern Hemisphere (ESO), K. Schwarzschild-str. 2,
              85748, Garching b. M\"unchen, Germany\\
              \email{fpatat@eso.org}
              \and 
              Dark Cosmology Centre, Niels Bohr Institute, University
              of Copenhagen, Juliane Maries Vej, DK-2100, Copenhagen, Denmark
              \and
               Istituto Nazionale di Astrofisica, Osservatorio Astronomico
               di Padova, v. Osservatorio n.5, I-35122, Padua, Italy
              \and
	      Queen's University - Belfast, BT7 1NN, Northern Ireland, UK
              \and
              Fundaci\'on Galileo Galilei-INAF, Telescopio Nazionale Galileo, 
              E-38700, Santa Cruz de la Palma, Tenerife, Spain
              \and
              Istituto Nazionale di Astrofisica, Osservatorio Astronomico
              di Catania, v. S. Sofia 78, I-95123, Catania, Italy
}

   \date{Received August 7, 2009; accepted October 2, 2009}

\abstract{}{}{}{}{} 
 
  \abstract
   {}
   {We study possible signs of asymmetry in
    the bright optical transient in NGC~300, to obtain
    independent information about the explosion mechanism, the progenitor
    star and its circumstellar environment.
   }
   {Using VLT-FORS1, we obtained low-resolution optical linear 
   spectropolarimetry of NGC~300~OT2008-1 on two epochs, 48 and 55 days after
   the discovery, covering the spectral range 3600--9330\AA.}
   {The data exhibit a continuum polarization at a very significant level. At 
   least two separate components are identified. The first is characterized 
   by both a strong wavelength dependency and a constant position angle 
   (68.6$\pm$0.3 degrees), which is parallel to the local spiral arm of 
   the host galaxy. The latter is aligned along a completely different position 
   angle (151.3$\pm$0.4). While the former is identified as arising in the 
   interstellar dust associated with NGC~300, the latter is most likely caused 
   by continuum polarization by dust scattering in the circumstellar 
   environment. 
   No line depolarization is detected in correspondence to the most intense 
   emission lines, disfavoring electron scattering as the source of intrinsic 
   polarization. This implies that there is a very small deviation from 
   symmetry in the continuum-forming region. Given the observed level of 
   intrinsic polarization, the transient must be surrounded by a significant 
   amount of dust 
   ($\geq$4$\times$10$^{-5}$ $M_\odot$), asymmetrically distributed within 
   a few thousand AU. This probably implies that one or more asymmetric 
   outflow episodes took place during the past history of the progenitor.} 
   {}


\authorrunning{F. Patat et al.}
\titlerunning{VLT spectropolarimetry of the optical transient in NGC~300}

   \maketitle
%

\section{\label{sec:intro}Introduction}

Many surveys have focused their interest on poorly known objects lying
in the gap in the absolute magnitude distribution that separates
luminous novae (M$_R \geq$ $-$10) and faint supernovae (M$_R \leq$
$-$14) (e.g., Rau et al. \cite{rau08}).  Although this gap had been
understood to be populated mostly by eruptions of luminous blue
variables (LBVs, see Maund et al. \cite{mau06}; Van Dyk et
al. \cite{van06}), it is now evident that new types of explosions may
be responsible for the transient events occurring in this luminosity
range, from exotic (LBV-like) outbursts of Wolf-Rayet stars
(Pastorello et al.  \cite{pasto07a}) to other long duration
transients, such as M85 OT2006-1 (Kulkarni et al. \cite{kul07}; Rau et
al. \cite{rau08}; Ofek et al.  \cite{ofek08}; Pastorello et
al. \cite{pasto07b}), whose nature remains debated. In this context, a
significant impulse to the study of objects in this gap was brought
about by the discoveries of two nearby objects, namely SN~2008S in
NGC~6946 and the 2008 luminous optical transient in NGC~300 (hereafter
NGC~300~OT2008-1).

NGC~300~OT2008-1 was discovered on May 14, 2008 (Monard
\cite{monard}). A spectrum obtained the day after, revealing
H$\alpha$, H$\beta$, Ca~II near-IR triplet, forbidden [Ca~II] doublet
in emission, and Ca~II H\&K absorptions (Bond, Walter \& Velasquez
\cite{bond08}), shared striking similarities with that of SN 2008S
(Prieto \cite{prieto08b}).  The photometric and spectroscopic
properties of the transient in NGC~300 led to the conclusion that this
was not a classical nova, an LBV, or a supernova (Bond et
al. \cite{bond09}). According to Bond et al. (\cite{bond08},
\cite{bond09}), the spectrum was reminiscent of that of V838 Mon
during the early phases of its outburst (Wisniewski et
al. \cite{wisniew}). Broad-band optical and near-IR photometric, plus
low and intermediate resolution spectroscopic follow-up observations
were presented by Bond et al. (\cite{bond09}), while Berger et
al. (\cite{berger09}) reported on UV, radio, and X-ray imaging and
high resolution spectroscopy.  Remarkably, no source was detected at
the position of NGC~300~OT2008-1 in the radio and X-ray observations,
which indicates that the transient was at least one order of magnitude
(in the radio domain) and two orders of magnitudes (in the X-rays
region) fainter than any SN discovered to date.  In addition,
high-resolution spectroscopy suggests the presence of a complex
circumstellar environment resulting either from previous asymmetric
ejections from the progenitor or from a companion wind (Bond et
al. \cite{bond09}; Berger et al. \cite{berger09}).

An inspection of pre-explosion, archival HST images revealed that no
star was detectable at the outburst position placing stringent upper
limits on the progenitor's luminosity, which was interpreted as an
additional argument in favor of a non-supernova origin (Berger \&
Soderberg \cite{berger08}). However, a progenitor candidate was
detected in pre-outburst Spitzer images (Prieto \cite{prieto08b};
Berger et al. \cite{berger09}), and it was identified as a massive
star (15-20 M$_\odot$), with a spectral energy distribution very
similar to that exhibited by the progenitor of SN~2008S (Prieto et
al. \cite{prieto08a}; Botticella et al. \cite{mt09}). By means of an
independent method based on the study of the surrounding stellar
population, a similar range for the progenitor mass (12-17 M$_\odot$)
was found by Gogarten et al. (\cite{gogarten}), while Bond et
al. (\cite{bond09}) gave a slightly lower mass range (10-15
M$_\odot$). An analysis of the Spitzer/IRS mid-IR spectrum of
NGC~300~OT2008-1 obtained three months after the outburst led Prieto
et al. (\cite{prieto09}) to conclude that, although the presence of a
massive star cannot be definitely excluded, a lower mass (6-10
M$_\odot$, carbon-rich AGB/super-AGB, or a post-AGB) precursor would
provide a better match to their mid-IR observations.

Albeit with some scatter in the estimate of the progenitor's mass,
most studies favor an exotic eruption of a moderate-to-massive star
for NGC 300 OT2008-1 (and SN 2008S). However, Botticella et
al. (\cite{mt09}) emphasized the overall similarity between the light
curves of these transients and those of type II-L SNe. The main
argument was the late-time flattening, which is consistent with the
slope expected from the radioactive decay of $^{56}$Co into
$^{56}$Fe. If this mechanism plays a role in powering the light curve
of these transients, then a SN explosion is the most likely engine
(electron-capture SNe, see Botticella et al. \cite{mt09}; Pumo et
al. 2009, submitted).

Soon after the discovery, we started an optical/near-IR follow-up
using a number of ground-based facilities. The results will be
presented and discussed in a forthcoming paper (Pastorello et
al. 2009, in preparation). In this paper we focus on the
spectropolarimetry of NGC~300~OT2008-1 that we obtained in July 2008
with VLT-FORS1 on two epochs separated by a week, with the aim of
detecting possible signs of asymmetries in the ejecta and/or in the
circumstellar environment that could provide additional constraints on
the progenitor's nature.

The paper is organized as follows. In Sect.~\ref{sec:obs} we discuss the
observations and the data reduction techniques. In Sect.~\ref{sec:flux} we 
present the flux spectra, while Sect.~\ref{sec:pol} deals with the
spectropolarimetric data sets. In Sect.~\ref{sec:disc}, we discuss the
results, and we summarize our conclusions in Sect.~\ref{sec:conc}.
Appendix~\ref{sec:app} contains an analysis of the effects of second
order contamination on spectropolarimetry and its application to the
data presented in this work, while Appendix~\ref{sec:appb} describes
the effect of multiple weak polarizers on the resulting polarization
signal.

\begin{figure}
\centerline{
\includegraphics[width=90mm]{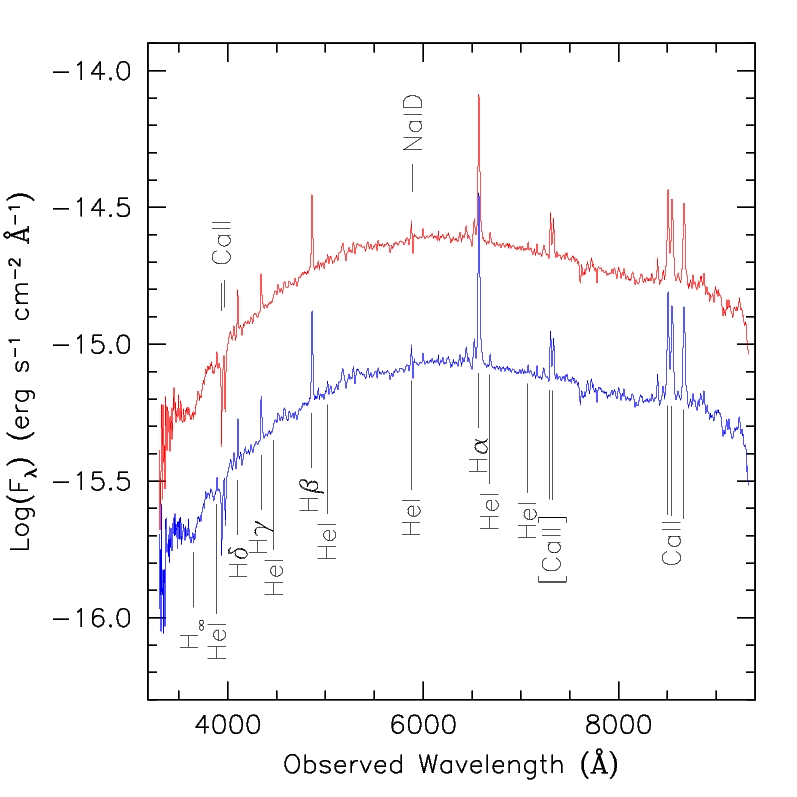}
}
\caption{\label{fig:compflux}Flux spectra of NGC~300~OT2008-1 on 01-07-2008
(upper) and 08-07-2008 (lower). For presentation, the spectrum of the
second epoch was shifted by $-$0.4. The main spectral features
are identified.}
\end{figure}

\section{\label{sec:obs}Observations and data reduction}

We observed the OT in NGC~300 on 2 different epochs, 48 and 55 days
after its discovery (Monard \cite{monard}), using the FOcal
Reducer/low-dispersion Spectrograph (hereafter FORS1), mounted at the
Cassegrain focus of the ESO--Kueyen 8.2m telescope (Appenzeller et
al. \cite{appenzeller}). In this multi-mode instrument, equipped with
a mosaic of two 2048$\times$4096 pixel (px) E2V CCDs, polarimetry is
performed by introducing into the optical path a Wollaston prism
(19$^{\prime\prime}$ throw) and a super-achromatic half-wave plate
(HWP). To reduce some known instrumental problems (see Patat \&
Romaniello \cite{patat06a}), we always used 4 half-wave plate angles
($\theta_i$=0, 22.5, 45 and 67.5 degrees). All spectra were obtained
with the low-resolution G300V grism coupled with a 1.0 arcsec slit,
giving a spectral range 3200-9330 \AA, a dispersion of $\sim$3.2 \AA\/
px$^{-1}$, and a resolution of 11.6 \AA\/ (FWHM) at 5800 \AA. To cover
the blue part of the optical spectrum ($\leq$4350\AA), where
significant Ca~II H\&K absorption is clearly evident, we have not used
an order sorting filter. Since the object is rather red ($B-V\sim$1.1,
see Sect.~\ref{sec:flux}), the second order contamination, which for
this grism starts at about 6600\AA\/ (O'Brien \cite{fors}), is
expected to be small and its effects on the spectropolarimetry of
NGC~300~OT2008-1 negligible. This is discussed more quantitatively in
Appendix~\ref{sec:app}.

\begin{figure*}
\centerline{
\includegraphics[width=100mm,angle=-90]{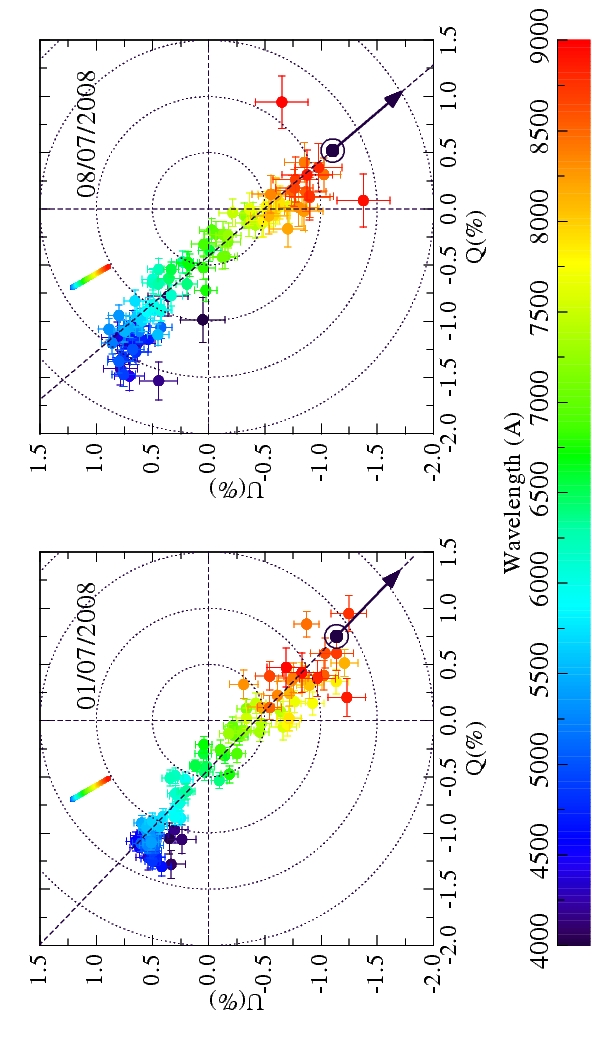}
}
\caption{\label{fig:quplot} Observed spectropolarimetry of
  NGC~300~OT2008-1 at the two epochs on the Stokes $Q-U$ plane. For
  presentation, the data were binned to 16 pixels (57\AA). The dotted
  circles are placed at constant polarization levels (0.5, 1.0, 1.5,
  and 2.0\%). The dashed line is a weighted least squares fit to the
  data in the displayed wavelength range. For comparison, the colored
  line sequence of dots in the upper left quadrant traces a Serkowski
  law (Serkowski, Matheson \& Ford \cite{serkowski}) for
  $\lambda_{max}$=4800\AA, $P_{max}$=1.4\%, and $K$=0.81 with an
  arbitrary position angle $\theta$=60 degrees. The circled dot
  indicates the position of $(Q_1,U_1)$ corresponding to the minimum
  polarization solution (see Sect.~\ref{sec:cont}), while the vector
  indicates the allowed region for $(Q_1,U_1)$.}
\end{figure*}

Data were bias and flat-field corrected, and wavelength calibrated
using standard tasks within IRAF\footnote{IRAF is distributed by the
  National Optical Astronomy Observatories, which are operated by the
  Association of Universities for Research in Astronomy, under
  contract with the National Science Foundation.}. The RMS error in
the wavelength calibration is about 0.7\AA.  The ordinary (upper) and
extraordinary (lower) beams were processed separately. Stokes
parameters, linear polarization degree, and position angle were
computed by means of specific routines written by us. Finally,
polarization bias correction and error estimates were performed
following the prescriptions described by Patat \& Romaniello
(\cite{patat06a}), while the HWP zeropoint angle chromatism was
corrected using tabulated data (O'Brien \cite{fors}). To increase the
signal-to-noise ratio, the final Stokes parameters were binned in
$\sim$25.8\AA\/ wide bins (8 pixels). This turns into an RMS error in
the polarization of 0.1\% at 6000\AA, where the continuum
signal-to-noise ratio per resolution element reaches its maximum (720
and 530 on the first and the second epoch, respectively). The spectra
are severely affected by fringing above $\sim$8000\AA, so that the
observed noise at the red edge of the wavelength range is
significantly greater than the statistical formal errors. The
instrument stability (including evolution of the fringing pattern) was
checked by comparing $\epsilon_Q$, $\epsilon_U$ pairs (Maund
\cite{maund08}) at the two epochs. Their average differences $\\angle
\Delta \epsilon\rangle= \langle \epsilon_Q - \epsilon_U\rangle$ were
found to be 0.07$\pm$ 0.02\% and 0.13$\pm$0.02\% on the first and the
second epoch, respectively. These values are within the measurement
errors per resolution element (typically $\leq$0.1\%), thus confirming
the stability of the instrument in the relevant time
interval. However, they also imply possible systematic errors of up to
about 0.1\%, which can be considered as the maximum accuracy one can
reach with FORS1 using 4 HWP positions (see also Patat \& Romaniello
\cite{patat06a}).

Flux calibration was achieved by observing a spectrophotometric
standard star and inserting full polarimetric optics (HWP angle set to
0). A log of the observations is given in Table~\ref{tab:log}.

\begin{table}
\caption{\label{tab:log}Log of the VLT-FORS1 observations of NGC~300~OT2008-1.}
\tabcolsep 3mm
\begin{tabular}{cccc}
\hline
 Date & MJD & Exp. Time & Airmass \\
 (UT) &     & (s)       &         \\
\hline
01-07-2008 & 54648.4 & 2$\times$500+2$\times$400 & 1.1\\
08-07-2008 & 54655.3 & 660+3$\times$500 & 1.2 \\
\hline
\end{tabular}
\end{table}

\section{\label{sec:flux}Flux spectra}

NGC~300~OT2008-1 was followed up in great detail by Bond et
al. (\cite{bond09}) and Berger et al. (\cite{berger09}), to which we
refer the reader for a comprehensive account of the spectroscopic
properties of this object and its evolution. In the following, we
provide a brief description of the flux spectra, while they will be
discussed in more detail in the analysis of the whole spectroscopic
data set of NGC~300~OT2008-1 (Pastorello et al., in preparation).

The spectra, obtained by coadding for each epoch the ordinary and
extraordinary beams for all the four HWP angles, are presented in
Fig.~\ref{fig:compflux}. Between the two epochs, which are separated
by seven days, no evolution is evident and the spectrum consists of a
red continuum with a number of superimposed narrow emission lines. As
Bond et al. (\cite{bond08}) pointed out, the most prominent spectral
lines can be identified with the Balmer series of Hydrogen, Ca~II
8498\AA, 8542\AA, 8662\AA, and [Ca~II] 7291, 7324\AA. The H lines can
be identified to wavelength as sort as H$\delta$ and the Balmer jump
at $\sim$3650\AA\/ is clearly detected
(Fig.~\ref{fig:compflux}). Additionally, our spectra appear to contain
He~I 3888\AA, 4571\AA, 5015\AA, 5876\AA, 6678\AA, 7065\AA\/ absorption
lines, which are clearly seen in the higher resolution spectra
presented and discussed by Berger et al. (\cite{berger09}). Finally,
Ca~II H\&K and Na~ID are visible in absorption and are contaminated by
unresolved features arising in the ISM (see also Berger et
al. \cite{berger09}).

\begin{figure}
\centerline{
\includegraphics[width=80mm]{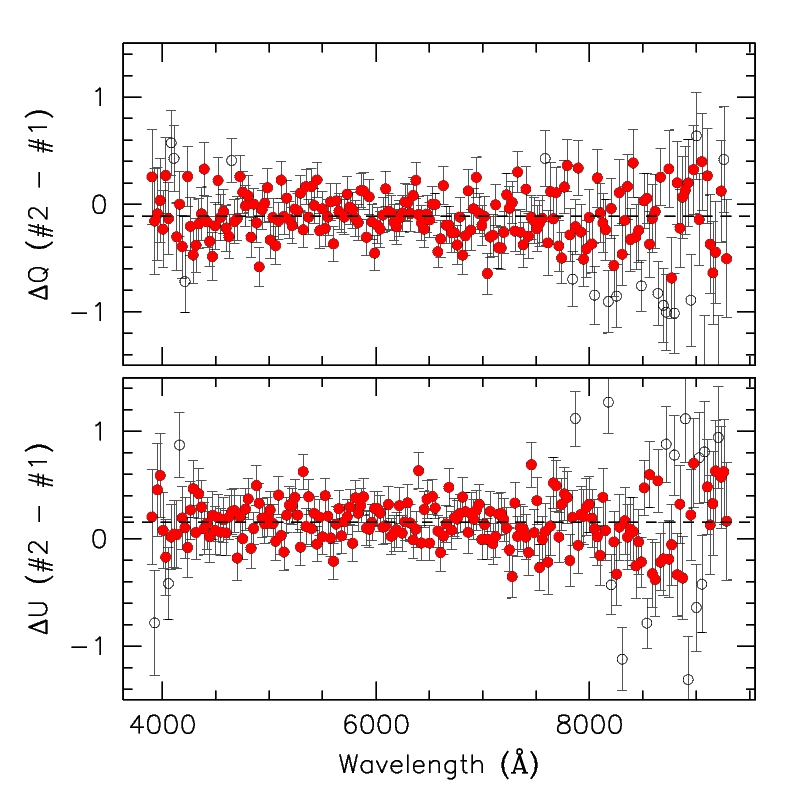}
}
\caption{\label{fig:qudiff} Differences in $Q$ (upper panel) and $U$ (lower 
panel) Stokes parameters between the two epochs (second minus
first). Empty symbols indicate the data points rejected by the
k-$\sigma$ clipping.}
\end{figure}

While the absolute flux scales should be compared to broadband
photometry when this becomes available, the overall shape of the
spectra is expected to be accurate to within 10-15\%.  The colors
deduced from synthetic photometry are $B-V$=+1.1, $V-R$=+0.6, and
$V-I$=+1.1 for both epochs, with RMS uncertainties of about 0.15 mag.
These values are fully compatible (within the quoted errors) with the
broadband photometry reported by Berger et al. (\cite{berger09}) for
June 19 ($B-V$=1.01$\pm$ 0.03; see their Table~2). Because of the
relatively high value of $B-V$, the flux contamination by the second
order above 6500\AA\/ is expected to be less than a few percent (see
Appendix~\ref{sec:app}).

The measured FWHMs of H$\alpha$, H$\beta$, and Ca~II 8662\AA\/ are
18.6$\pm$0.3\AA, 13.8$\pm$0.2\AA\/ and 23.8$\pm$0.4\AA, respectively.
Once corrected for the instrumental resolution (11.6\AA), these values
correspond to expansion velocities of 660, 460, and 720 km
s$^{-1}$. In general, the spectrum of NGC~300~OT2008-1 is very similar
to those reported for SN~2008S by Smith et al. (\cite{smith}). Based
on the very low luminosity and expansion velocities, these authors
concluded that the object previously designated as SN~2008S is
actually a SN impostor (Van Dyk et al. \cite{vandyk}), similar to the
giant eruption of a luminous blue variable, and shares the spectral
properties of the Galactic hypergiant IRC+10420 (Jones et
al. \cite{jones}). Here we note that, in general, SN~2008S differs
from a classical impostor, in that SN~2008S has a linear and slow
evolving luminosity decline, a spectrum that matches those of LBV
eruptions well, excluding its very strong [Ca~II] doublet. Based
on similar considerations, both Bond et al. (\cite{bond09}) and Berger
et al. (\cite{berger09}) concluded that NGC~300~OT2008-1 was the
eruption of a dust-enshrouded, massive star. An alternative scenario
was proposed by Prieto et al. (\cite{prieto08a}), Thompson et
al. (\cite{thompson}), and Botticella et al. (\cite{mt09}), who argued
that SN~2008S and NGC~300~OT2008-1 were generated by the explosion of an
electron-capture SN.

\begin{figure}
\centerline{
\includegraphics[width=80mm]{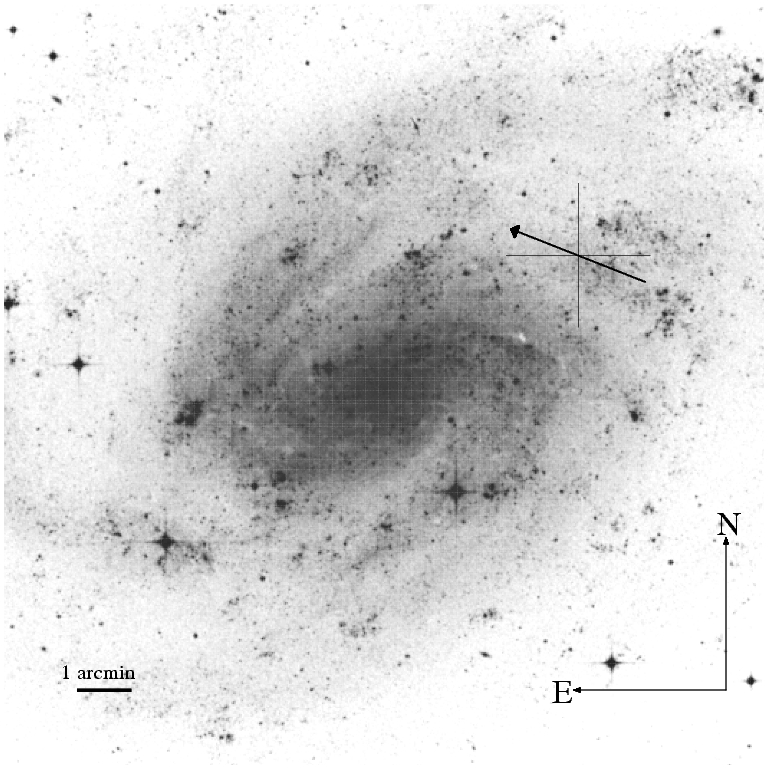}
}
\caption{\label{fig:fchart} A DSS image of NGC~300. The location of
  the optical transient is marked by a cross. The arrow indicates the
  polarization position angle of the dominant axis (the average of the
  values measured during the two epochs was used: $\theta$=68.6
  degrees).}
\end{figure}

\section{\label{sec:pol}Spectropolarimetry}

The spectropolarimetric data for NGC~300~OT2008-1 are presented in the
$Q-U$ plane in Fig.~\ref{fig:quplot}. The object exhibits a
significant polarization, which reaches a maximum level of about
1.4\%. During both epochs, the data cover a range of polarization
values which are definitely much higher than expected for a typical
Serkowski law (Serkowski et al. \cite{serkowski}; see also next
section) with a similar maximum polarization (see
Fig.~\ref{fig:quplot} and its caption). No obvious change in
polarization is seen across the most prominent emission lines
(Sect.~\ref{sec:lines}).

\subsection{\label{sec:cont}Continuum polarization}

The most remarkable aspect of the $Q-U$ plot is the alignment of the
data points along a straight line, which is indicative of a
polarization component with a clearly defined dominant axis. This does
not pass through the origin of the $Q-U$ plane, implying that there
must be an additional component with a polarization wavelength
dependence which differs from that along the dominant axis (see
Appendix~\ref{sec:appb}). We also note that the two data sets span
similar ranges of values along the dominant direction, but that a
rigid shift is present between the two epochs. This is more clearly
illustrated in Fig.~\ref{fig:qudiff}, where we plot the differences
$\Delta Q$ and $\Delta U$ between the two epochs. The clipped average
differences (computed in the wavelength range 3900-9300\AA) are
$\langle \Delta Q\rangle$=$-$0.11$\pm$0.02\% (191 spectral bins) and
$\langle \Delta U\rangle$=+0.14$\pm$0.02\% (193 spectral bins), and
the RMS deviation from the average is 0.23 and 0.21\% for the two
Stokes parameters, respectively. Given that the observed grey
variation is very small and comparable to the instrument stability
between the two epochs ($\sim$0.1\%, see Sect.~\ref{sec:obs}), it is
unclear whether this is real or caused by an incompletely removed
instrumental effect.

\begin{figure}
\centerline{
\includegraphics[width=80mm]{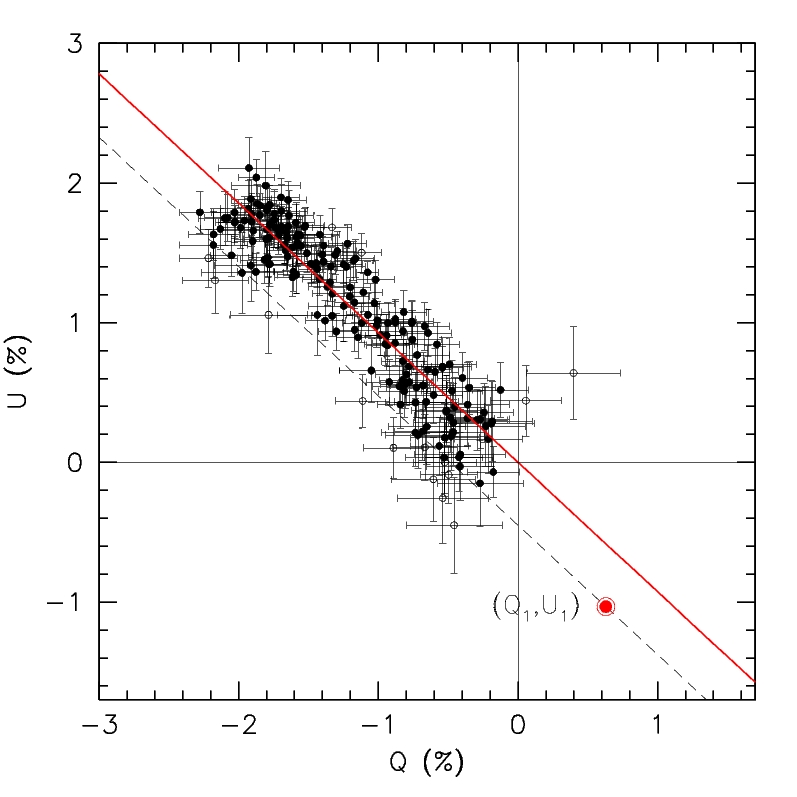}
}
\caption{\label{fig:quplane1} Minimum polarization solution for 08-07-2008.
The dashed line traces the observed dominant axis, while the point
$(Q_1,U_1)$ indicates the wavelength-independent component. The empty
circles mark the rejected data points.}
\end{figure}

To characterize the dominant axis, we fitted the data with a
linear relation

\begin{equation}
\label{eq:dominant}
U=U_0 \; + Q \; \tan 2\theta
\end{equation}

\noindent The results of the least squares fitting for the two epochs
in the wavelength range 4000-9000\AA\/ are reported in
Table~\ref{tab:dominant} (upper part), which includes the RMS
deviation from the best-fit relation ($\sigma$), the Pearsons linear
correlation coefficient ($r_{xy}$) and $\chi^2$ per degrees of freedom
(DOF). To exclude the most deviant data points, we performed a
preliminary fit to all data, applied a 1.7-sigma rejection and finally
refit the remaining data. The large correlation factors confirm the
presence of a dominant axis at a high significance level.  This
indicates that the main source of polarization is interstellar
extinction. Interstellar polarization (ISP), caused by the presence of
aligned asymmetric dust grains along the line of sight, is
characterized by a wavelength-independent position angle (Serkowski et
al. \cite{serkowski}; Whittet et al. \cite{whittet}).  This conclusion
is supported by the observed dominant axis being tangential to the
spiral arm of NGC~300 close to the outburst position, as shown in
Fig.~\ref{fig:fchart}. Since dust grains tend to be aligned in the
direction of the galactic magnetic field, which in turn follows the
spiral pattern (see Scarrot, Ward-Thompson \& Warren-Smith
\cite{scarrot}), this strongly indicates that not only the observed
polarization is caused by directional extinction, but also that the
polarization arises within interstellar material associated with a
spiral arm of NGC~300. Similar alignments were detected using
spectropolarimetry in two supernovae, namely SN~2001el (Wang et
al. \cite{wang03}) and SN~2006X (Patat et al. \cite{patat09}). From
now on we therefore assume that this component is constant in time.

The estimated position angles at the two epochs differ by 2.0$\pm$0.5
degrees. Since this variation is only marginally significant, we adopt
the average of the two values (68.6$\pm$0.5 degrees) as the
representative value. For consistency, we rotated the two data sets by
$-$1.0 and +1.0 degrees, respectively. The results of the least
squares fitting after this operation are shown in
Table~\ref{tab:dominant} (lower part).  The recomputed average
differences in $Q$ and $U$ are $\langle\Delta
Q\rangle$=$-$0.12$\pm$0.02\% and $\langle \Delta
U\rangle$=+0.11$\pm$0.02\%. This means that the systematic,
wavelength-independent shift observed between the two epochs is not
related to this small rotation.

\begin{table}
\tabcolsep 1.8mm
\caption{\label{tab:dominant}Dominant axis least squares fit parameters.}
\begin{tabular}{cccccc}
\hline
Date  & $U_0$ & $\theta$ & $\sigma$ & $r_{xy}$ & $\chi^2$/DOF\\
      & (\%)  & (deg)    & (\%)     &          &              \\
\hline
01-07-2008 & $-$0.42 (0.02) & 69.6 (0.4) & 0.19 & $-$0.94 & 432/180\\
08-07-2008 & $-$0.46 (0.02) & 67.6 (0.5) & 0.23 & $-$0.92 & 333/179\\
\hline
01-07-2008 & $-$0.43 (0.02) & 68.6 (0.4) & 0.20 & $-$0.94 & 463/180\\
08-07-2008 & $-$0.45 (0.02) & 68.6 (0.5) & 0.22 & $-$0.92 & 311/179\\
\hline
\multicolumn{6}{l}{Note: Values in parenthesis are the statistical RMS uncertainties.}\\
\multicolumn{6}{l}{The lower part of the table reports the same values after rotation}\\
\multicolumn{6}{l}{to the average dominant axis ($\theta$=68.6 degrees).}
\end{tabular}
\end{table}

As we said, the position of the points on the $Q-U$ plane requires an
additional polarization component. The simplest solution is given by a
wavelength-independent polarization, represented by one single point
$(Q_1,U_1)$ on the $Q-U$ plane. In these circumstances, the total,
observed Stokes parameters $Q$ and $U$ are given by $Q = Q_1 +
Q_2(\lambda)$ and $U = U_1 + U_2(\lambda)$, where $Q_2$ and $U_2$ are
the Stokes parameters that characterize the polarization aligned along
the dominant axis\footnote{In the weak polarization approximation the
  components are vectorially additive. See
  Appendix~\ref{sec:appb}.}. Of course, if the ratio of $Q_2$ to $U_2$
must remain constant (as required by the wavelength-independent
position angle), $Q_1$ and $U_1$ must obey Eq.~\ref{eq:dominant},
i.e. $(Q_1,U_1)$ must be aligned along the {\it observed} dominant
axis. One additional constraint related to the definitions of $Q$ and
$U$ is that all the $(Q_2,U_2)$ points have to be confined to one
single quadrant of the $Q-U$ plane (Wang et al. \cite{wang03}), which,
in our case, is that with $Q<0$ and $U>0$. This imposes that $Q_1$ is
higher than a certain value $Q_{min}$, but does not prevent $Q_1$
being any higher arbitrary value (see the arrow in
Fig.~\ref{fig:quplot}). For this reason, although constrained, the
solution is not unique. As a consequence, while the position angle of
the dominant axis ($\theta_2$) is fixed, the corresponding
polarization $P_2(\lambda)=\sqrt{Q_2^2(\lambda)+U_2^2(\lambda)}$ is
undefined up to an additive constant, which depends on the particular
choice of the wavelength-independent polarization
$P_1=\sqrt{Q_1^2+U_1^2}$. Since $(Q_1,U_1)$ must lie along the
dominant axis, the corresponding value of $\theta_1=\arctan U_1/Q_1$
is allowed to vary only within a rather limited range (smaller than 10
degrees), regardless how large the value of $Q_1$ is.

\begin{figure}
\centerline{
\includegraphics[width=80mm]{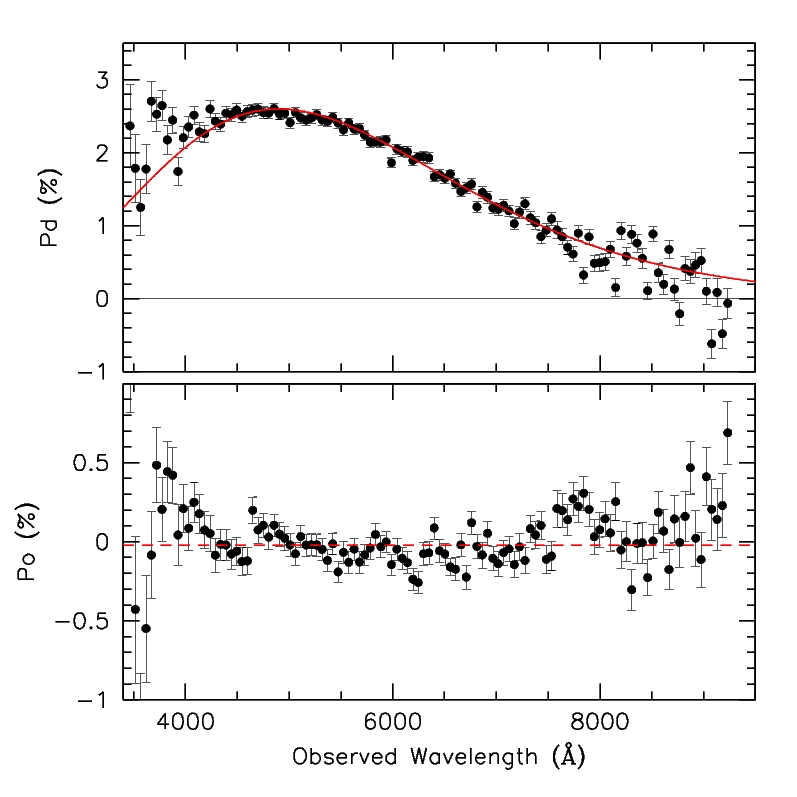}
}
\caption{\label{fig:rotate1} Dominant (upper panel) and orthogonal 
(lower panel) polarization on 01-07-2008 after subtracting the
wavelength-independent component $\vec{P_1}$.  For presentation, the
data were binned to 16 pixel (57\AA). The solid curve in the
upper panel is a Serkowski law with $P_{max}$=4.8\%,
$\lambda_{max}$=4250\AA, and $K$=3.0.}
\end{figure}

\subsubsection{\label{sec:msp}Minimum polarization solution}

In the following, we consider what we will indicate as minimum
polarization solution (MPS), i.e., the solution one obtains for
$Q_1=Q_{min}$, and that leads to the minimum possible values of the
polarization degree for both components. Given the wavelength
dependency shown by the data (Fig.~\ref{fig:quplot}), this means that
the MPS corresponds to the case where the polarization is close to
zero at the red edge of the observed wavelength range ($\sim$9300\AA).
The solution was found as the value of $Q_1$ that produces the minimum
constant value of $P_2(\lambda)$ at any given wavelength, while
fulfilling the constraints described above. Since we had shown that
the two data-sets differ by a wavelength-independent quantity
(cf. Fig.~\ref{fig:qudiff}), we computed $Q_1$ for the second
epoch and set the value for the first epoch such that $\Delta
Q_1$=$-$0.12\% (see above). Within the errors, this ensures that
$P_2(\lambda)$ does not change between the two dates. The results of
this procedure are presented in Table~\ref{tab:sol} and illustrated in
Fig.~\ref{fig:quplane1} for the second epoch. What emerges from this
analysis is that the additive, wavelength-independent component is
polarized at a highly significant level, which reaches at least
1.2\%. Its position angle (151.3$\pm$0.4 degrees, averaged over the two
epochs) is very different from that of the dominant component,
indicating that they must arise from dissimilar sources
(see the discussion in Sect.~\ref{sec:disc}).

\begin{table}
\tabcolsep 2.4mm
\caption{\label{tab:sol}Minimum polarization solutions for the two epochs.}
\begin{tabular}{ccccc}
\hline
Date       & $Q_1$ & $U_1$ & $P_1$ & $\theta_1$ \\
           & (\%)  & (\%)  & (\%)  & (deg)      \\
\hline
01-07-2008 & +0.75 & $-$1.12(0.03) & 1.35(0.03) & 151.9(0.5)\\
08-07-2008 & +0.63 & $-$1.03(0.03) & 1.21(0.03) & 150.7(0.5)\\
\hline
\multicolumn{5}{l}{Values in parenthesis are the statistical RMS 
uncertainties.}
\end{tabular}
\end{table}

When, as in this case, there is a marked dominant direction, it is
useful to compute the components parallel ($P_d$) and orthogonal
($P_o$) to this axis, as suggested by Wang et al. (\cite{wang03}).
This is achieved by the following roto-translation in the $Q-U$
plane:

\begin{displaymath}
\begin{array}{ccc}
P_d &=&  (Q-Q_1) \cos 2\theta_2 + (U-U_1) \sin 2\theta_2, \\
P_o &=&  (U-U_1) \cos 2\theta_2 - (Q-Q_1) \sin 2\theta_2. \\
\end{array}
\end{displaymath}

\begin{figure}
\centerline{
\includegraphics[width=80mm]{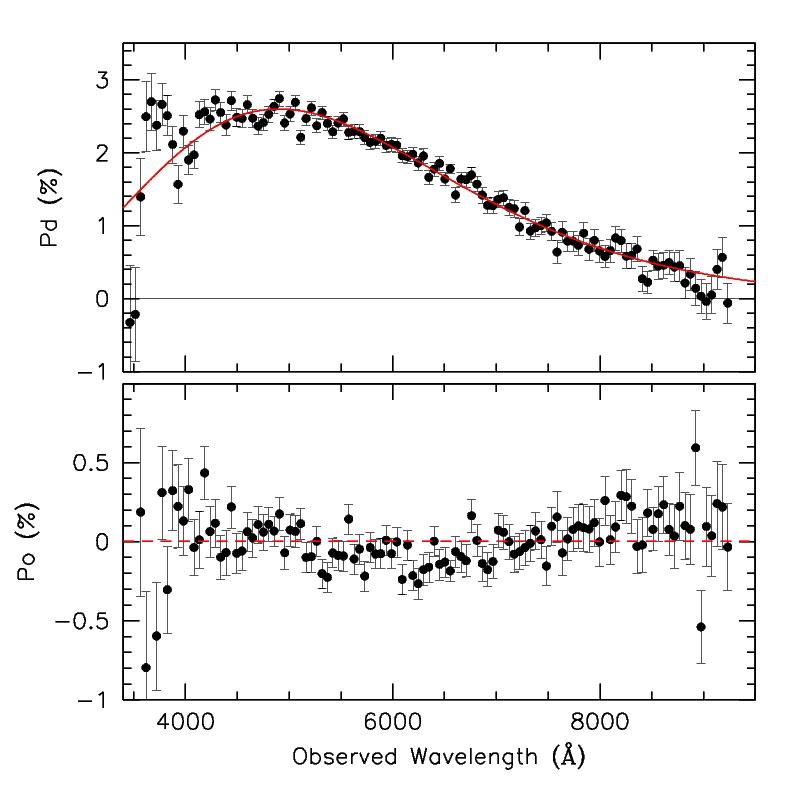}
}
\caption{\label{fig:rotate2} Same as Fig.~\ref{fig:rotate1} for 08-07-2008.}
\end{figure}

\noindent While $P_d$ is undefined up to an additive constant
(dependent on the particular choice of $Q_1$), $P_o$ is not. Indeed,
$P_o$ is purely related to the deviations from the dominant axis. The
decomposition is shown in Figs.~\ref{fig:rotate1} and
\ref{fig:rotate2} for the two epochs. Two important results emerge from
these plots: i) $P_d$ has a smooth wavelength dependence, reaching
a maximum (2.6$\pm$0.1\%) at about 4900\AA\/ and decreasing on
either sides of the peak much faster than a Galactic Serkowski law
(Serkowski et al. \cite{serkowski}); ii) $P_o$ is consistent with a
null value. The RMS deviations in the wavelength range 4000-9000\AA\/
are 0.13\% and 0.16\% during the two epochs, respectively.  These values
are fully compatible with the estimated measurement errors, and thus
the two components $\vec{P_1}$ and $\vec{P_2}$ are sufficient to
explain all the observed variance in the residuals. During the second
epoch, there might be an indication of a mild, smooth wavelength
dependency in $P_o$ (see Fig.~\ref{fig:rotate2}, lower panel), but the
statistical significance appears to be marginal. The same is true for
a small bump seen at about 7800\AA\/ during the first epoch
(Fig.~\ref{fig:rotate1}, lower panel).

For highly reddened Galactic stars, Serkowski et
al. (\cite{serkowski}) showed that the linear polarization degree
changes with wavelength according to the empirical law

\begin{displaymath}
P(\lambda)=P_{max} \; \exp\left ( -K\; \ln^2 \; 
\frac{\lambda_{max}}{\lambda}\right ),
\end{displaymath}

\noindent where $\lambda_{max}$ is the wavelength that corresponds to
the maximum polarization $P_{max}$. Witthet et al. (\cite{whittet})
also demonstrated that $K=0.01 + 1.66\; \lambda_{max}$, where
$\lambda_{max}$ is expressed in microns. Typical values of $K$ range
from 0.7 to 1.4. A comparison between the NGC~300~OT2008-1 data and
the behavior expected on the basis of the Serkowski law immediately
shows that the observed wavelength dependence is much steeper
(Fig.~\ref{fig:quplot}), indicative of values of $K$ higher than 5
(see Figs.~\ref{fig:rotate1}, \ref{fig:rotate2}). For comparison,
Winsniewski et al. (\cite{wisniew}) report $K\simeq$1 for the
Galactic, peculiar variable V838 Monocerotis, for which they could
reproduce the observational data with a Serkowski law (see their
Fig.~5). Interestingly, $P_d$ is reasonably well fitted by a Serkowski
law with $P_{max}$=2.6$\pm$0.1\%, $\lambda_{max}$=4880$\pm$40\AA, and
$K$=5.6$\pm$0.1 (the Witthet et al. relation gives $K$=0.82 for the
observed $\lambda_{max}$).

This marked deviation indicates that the dust mixture responsible for
the polarization most likely differs from a typical Galactic
mixture. Therefore, Serkowski's Galactic relation between color excess
and maximum polarization ($P_{max}(\%)\geq 9\; E_{B-V}$) probably does
not hold for NGC~300~OT2008-1. In this respect, we note that the ISP
wavelength dependency observed in several supernovae also clearly
deviates from the Galactic Serkowski law (Leonard \& Filippenko
\cite{leonard}; Leonard et al. \cite{leonard02}; Maund et al.
\cite{maund}; Patat et al. \cite{patat09}). For the hypothesis that
the dominant polarization detected in NGC~300~OT2008-1 is produced by
dust extinction, as all elements seem to indicate, one can attempt to
estimate reddening from the observed level of polarization. Blindly
using $P_{max}$=2.6\% in the quoted relation, one obtains an upper
limit to the color excess given by $E_{B-V}\leq$0.3. The observed
deviation from a Galactic Serkowski law casts serious doubts on the
reliability of this estimate, and what we quote here is just for the
sake of completeness. To our knowledge no polarimetric study on
NGC~300 exists in the literature. Therefore, it is not possible to
tell whether the observed wavelength dependency is typical of the host
galaxy or is a peculiarity of the line of sight to NGC~300~OT2008-1.
However, we note that the extinction law along the line of sight to
young clusters in NGC~300 was reported to be highly variable (Roussel
et al. \cite{roussel}).

Based on photometric considerations, Bond et al. (\cite{bond09})
report $E_{B-V}$=0.4, while Berger et al. (\cite{berger09}) derive
$E_{B-V}$=0.05$\pm$0.05 from the equivalent width of the Na~I D
lines. They conclude that this low value for the reddening is
consistent with the lack of obvious evidence of extinction in their
optical and UV spectra, and point out that this is in line with
evidence of dust destruction in the immediate surroundings of the
outbursting star. The extinction arising within the Milky Way is very
small ($E_{B-V}$=0.013, Schlegel, Finkbeiner \& Davis \cite{schlegel})
and the bulk of reddening has to occur within the host galaxy, and so
is most likely the case for interstellar polarization. The compilation
of Heiles (\cite{heiles}) contains a few stars close in galactic
coordinates to
NGC~300\footnote{http://vizier.cfa.harvard.edu/viz-bin/VizieR?-source=II/226}.
All of them have linear polarizations $\leq$0.1\%, in close agreement
with the low Galactic extinction along this line of sight.

\begin{figure}
\centerline{
\includegraphics[width=80mm]{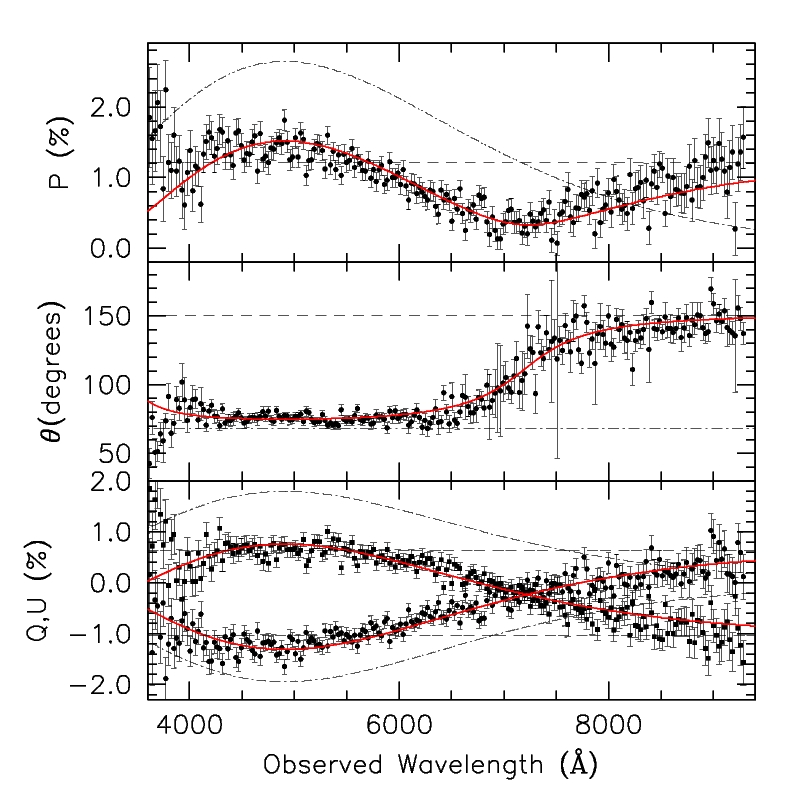}
}
\caption{\label{fig:globsol2}Best fit solution for 08-07-2008 
compared to the observed Stokes parameters (lower panel), position
angle (upper panel) and polarization level (upper panel).  The curves
trace the wavelength-independent component $\vec{P_1}$ (dashed), $\vec{P_2}$
(dotted-dashed) and the global solution (solid).}
\end{figure}

The best-fit global solution is compared to the observed Stokes
parameters and position angle in Fig.~\ref{fig:globsol2} for the data
acquired on 2008 July 8. The fit is fairly good for this epoch
($\chi^2$=1.1, RMS=0.20\%), while for the first epoch the observed
data points show higher scatter redwards of $\sim$7000\AA\/
($\chi^2$=2.8, RMS=0.28\%).

The synthetic $B,V,R$, and $I$ broad-band polarizations are reported
in Table~\ref{tab:broad}. As we have shown for the spectra, the
observed variations (all highly significant) can be explained in terms
of evolution in the wavelength-independent component. After applying
the systematic differences $\langle \Delta Q\rangle$ and $\langle
\Delta U\rangle$ to the second epoch and re-computing the synthetic
broad-band polarimetry, both the polarization angle and degree are
consistent (within the errors) with those derived for the first epoch
(Table~\ref{tab:broad}, last row).  Interestingly, the variation in
$Q_1$ and $U_1$ takes place along a direction (PA=154.1$\pm$3.2)
consistent with the average position angle of $\vec{P_1}$
(151.3$\pm$0.4). If this variation were real, this would imply that
the wavelength-independent component evolves just within its
polarization level, while the position angle remains practically
unchanged.  This change in the polarization level is indicative of an
intrinsic origin. However, since the observed variation is within the
typical instrumental stability limits ($\sim$0.1\%, see
Sect.~\ref{sec:obs}), no solid conclusion about its nature can be
reached.

\begin{table}
\tabcolsep 1.2mm
\caption{\label{tab:broad}Synthetic broadband polarimetry of NGC~300~OT2008-1.}
\begin{tabular}{lcccccccc}
\hline
Epoch&\multicolumn{2}{c}{B}&\multicolumn{2}{c}{V}&\multicolumn{2}{c}{R}&\multicolumn{2}{c}{I}\\ 
     & P (\%) & $\theta$   & P (\%) & $\theta$  & P (\%) & $\theta$ & P (\%) & $\theta$  \\
\hline
\#1  & 1.21 & 79.0 & 1.04 & 77.5 & 0.55 & 84.6 & 0.67 & 141.7 \\
\#2  & 1.36 & 76.4 & 1.18 & 74.1 & 0.71 & 78.2 & 0.53 & 133.8 \\
     &      &      &      &      &      &      &      &       \\
\#2(*)& 1.21 & 78.4 & 1.02 & 76.1 & 0.55 & 82.3 & 0.67 & 139.4 \\
\hline
\multicolumn{9}{l}{(*) Recomputed after applying $<\Delta Q>$=$-$0.11,$<\Delta U>$=+0.14.}\\
\multicolumn{9}{l}{Note: The RMS errors on polarization levels and position angles are}\\
\multicolumn{9}{l}{0.02\% and 0.3 degrees, respectively.}
\end{tabular}
\end{table}

\subsection{\label{sec:lines}Line polarization}

Although the resolution of our spectra is rather low ($\sim$12\AA\/
FWHM), we inspected the data searching for possible line effects
corresponding to the most prominent emission lines. In doing this,
we first vectorially subtracted the contribution of the component
$\vec{P_2}$ (which we assumed to be of interstellar origin and to be
constant across the line profile) directly estimated fitting the
adjacent $Q_2(\lambda)$ and $U_2(\lambda)$ data points and
interpolating to the line wavelength\footnote{Note that this is a pure
  shift of $Q$ and $U$ that does not modify their shape.}. The case of
H$\alpha$, the most intense feature (see Fig.~\ref{fig:compflux}), is
presented in Fig.~\ref{fig:hal1} for the first epoch. Across the line
profile there is no evidence of the polarization changes that were
previously detected, for instance, in Nova Cygni 1992 (Bjorkman et
al. \cite{bjork}), Nova Scuti 2003 (Kawabata et al. \cite{kawabata}),
and V838 Mon (Wisniewski et al. \cite{wisniew}). A similar behavior is
observed during the second epoch and for all other prominent emission
lines (H$\beta$ and Ca~II NIR triplet), in the sense that the observed
polarization remains at the wavelength-independent level of
$\vec{P_1}$.  If the emission line is essentially unpolarized
(Harrington \& Collins \cite{harrington}), the ratio of line to
continuum polarization is given by the following semi-empirical
relation (McLean \& Clarke \cite{mclean}):

\begin{displaymath}
\frac{P(\lambda)}{P_c} = \frac{1}{1+F(\lambda)/F_c},
\end{displaymath}

\noindent where $F(\lambda)$ is the line intensity profile and $F_c$
is the adjacent continuum level. Based on this relation and given
that, at the peak of H$\alpha$, $F(\lambda)/F_c\simeq$3
(Fig.~\ref{fig:hal1}), one would predict the level of polarization to
drop to $\sim$0.25 $P_c$ at the line center. The relatively high
signal-to-noise ratio achieved at the H$\alpha$ line peak ($\sim$700
in the central, 3 pixel bin) allowed us to place an upper limit on the
depolarization, which must be smaller than 0.2\% during both
epochs. This definitely excludes any depolarization at the level
predicted by the above relation.

Finally, we searched for a polarization signal associated with the
pronounced Ca~II H\&K absorption, after vectorially subtracting the
adjacent continuum polarization. Although we achieved a marginal
detection during the first epoch (0.6$\pm$0.2\%), this signal is not
present during the second and, therefore, we do not attach too great
significance to this finding.

\section{\label{sec:disc}Discussion}

The data presented in this paper show that the polarization observed
in NGC~300~OT2008-1 is generated by at least two, almost
perpendicular, components $\vec{P_1}$ and $\vec{P_2}$. Because of the
constant polarization angle (rather precisely aligned along the local
spiral arm of NGC~300), and the wavelength dependence (reminiscent of
a Serkowski law), we have interpreted $\vec{P_2}$ as being caused by
ISP. The unusual shape of $P_2(\lambda)$, with its large slope
parameter $K$, might be indicative of a dust mixture that differs
in both chemical composition and grain size from that typical of the
Milky Way. As of today, there is no consistent model that explains the
empirical Serkowski law (Draine \cite{draine03a}), and it is therefore
impossible to derive the dust properties from the observed
wavelength dependence.

After reasonably constraining the origin of $\vec{P_2}$, what remains
to be clarified is the nature of $\vec{P_1}$.

\subsection{\label{sec:electron}Electron scattering}

The most natural source of wavelength-independent linear polarization
in a stellar envelope is Thomson scattering by free electrons (van den
Hulst \cite{vdhulst}). This is, for instance, the source of continuum
polarization in supernovae (H\"oflich \cite{hoeflich}). More
generally, polarization by electron scattering is supposed to
originate in stars with ionized and spatially extended atmospheres,
such as Be and shell stars, Of and Wolf-Rayet stars,and early-type
supergiants (Brown \& McLean \cite{brown}). A non-null net
polarization is indicative of an asymmetry in the continuum emitting
region, and can be understood in terms of partial cancellation of the
intrinsic polarization generated by the Thomson scattering. In the
case of NGC~300~OT2008-1, and based on the assumption of a spheroidal
photosphere, the measured level of polarization ($P_1\geq$1.2\%) would
indicate a substantial asphericity, of axial ratio $\leq$0.8
(H\"oflich \cite{hoeflich}). This geometry would produce a
wavelength-independent polarization, characterized by a constant
polarization angle (perpendicular to the major axis), similar to what
is observed.

Evidences of electron scattering polarization (at a level of about
0.5\%) were found in the early phases of Nova Cygni 1992 (Bjorkman et
al. \cite{bjork}) and V838 Mon (Wisniewski et al. \cite{wisniew}). In
both cases this was interpreted in terms of asymmetry in the
continuum-forming region, identified with a flattened, spheroidal shell
ejected during the outburst. Since in the optically thin case the
polarization level is proportional to the electron-scattering optical
depth (Brown \& McLean \cite{brown}), the observed decrease in the
polarization level is produced by the expansion of the shell. The
systematic, gray-decline that we observe between our two epochs (see
Sect.~\ref{sec:cont}) might be interpreted along the same lines.

An important problem with the electron scattering scenario in
NGC~300~OT2008-1 is the absence of line depolarization
(Sect.~\ref{sec:lines}). This effect is expected when the emission
lines arise in an outer ionized region, where Thomson scattering has a
lower optical depth (Harrington \& Collins \cite{harrington}; Clarke
\& McLean \cite{clarke}). It is observed in a variety of objects,
including Be stars (McLean \& Clarke \cite{mclean}), novae (Bjorkman
et al. \cite{bjork}; Kawabata et al. \cite{kawabata}), supernovae
(SN~1987A, Cropper et al. \cite{cropper}), eruptive events such as
V838 Mon (Wisniewski et al. \cite{wisniew}), and the post-red
supergiant IRC+10420 (Patel et al. \cite{patel}). However, we note
that there are exceptions, in which continuum polarization is observed
but no line effect is detected, both in Be stars (McLean \& Clarke
\cite{mclean}; Quirrenbach et al. \cite{quirrenbach}) and in Nova
Sagittarii 1999 (Kawabata et al. \cite{kawabata00}).

\begin{figure}
\centerline{
\includegraphics[width=80mm]{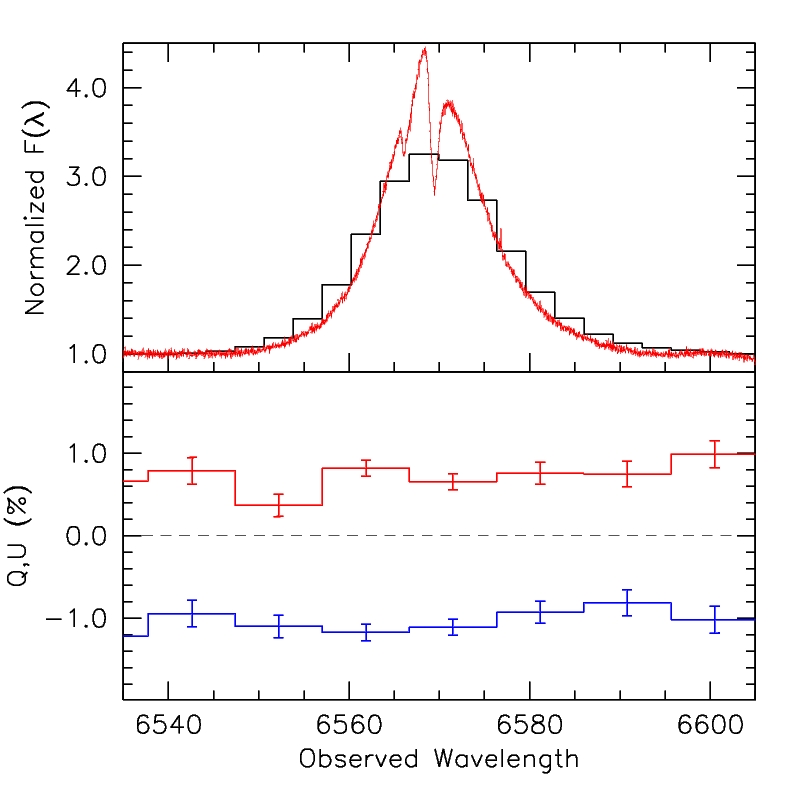}
}
\caption{\label{fig:hal1} Spectropolarimetry of the H$\alpha$ line on 
01-07-2008.  The data were binned to 3 pixel (9.6\AA). The flux
spectrum in the upper panel is unbinned. The thin colored line traces
a high-resolution spectrum ($\Delta \lambda/\lambda\sim$48,000)
obtained with VLT-UVES on 02-07-2008 (Pastorello et al. in
preparation).}
\end{figure}

The narrow absorption profile detected on top of the emission lines
(Bond et al. \cite{bond09}; Berger et al. \cite{berger09}. For the
case of H$\alpha$ see also Fig.~\ref{fig:hal1} here, upper panel)
might enhance the polarization within the overall line profile. A
similar phenomenon has been proposed to explain the absence of line
depolarization in the shell star $\eta$ Cen (McLean \& Clarke
\cite{mclean}). Given the modest depth of the narrow absorption, we
however propose this is not sufficient to explain the absence of
depolarization. Insufficient resolution can also be the responsible
for the lack of detection of line effects. However, we tend to exclude
this for our data, since the lines are partially resolved.

Another possibility is that the continuum and the emission lines in
NGC~300~OT2008-1 arise in similar regions, i.e., that the lines do not
form very far above the pseudo-photosphere, so that both the continuum
and the lines are subject to roughly the same amount of electron
scattering. Although this scenario can only be verified through NLTE
calculations, we note that this is quite unlikely. There is indeed
strong evidence that in both SN~2008S and in NGC~300~OT2008-1 lines
and continuum originate in very different regions. While the photospheric
temperature and radius decrease with time, the emission lines evolve
very slowly (Berger et al. \cite{berger09}; Botticella et
al. \cite{mt09}). This and the very high Balmer decrement lead to the
conclusion that the emission lines are generated by the interaction
between the ejected shell and the pre-existing circumstellar material,
while the continuum forms in an inner region (Berger et
al. \cite{berger09}).

Finally, there is a third possibility, i.e., that the polarization
source is located outside both the photosphere and the line-forming
region. This would probably exclude Thomson scattering by free
electrons as the source of the observed polarization and requires an
alternative explanation (see next section).

The absence of electron scattering in NGC~300~OT2008-1 can be
explained in terms of a spherically symmetric photosphere. Although
NGC~300~OT2008-1 and IRC+10420 have similar spectroscopic properties
(Smith et al. \cite{smith}), the post-red hypergiant has a continuum
polarization that is higher than 2\%, which is indicative of a
significative asymmetry in the continuum-forming region (Patel et
al. \cite{patel}).  Whatever the reason for the lack of
line-depolarization in NGC~300~OT2008-1, it definitely indicates a
radical difference between the physical conditions in the
continuum/line-forming regions of the two objects.

\subsection{\label{sec:dust}Dust scattering}

To explain the absence of line effects in Nova Sagittarii 1999,
Kawabata et al. (\cite{kawabata00}) proposed that continuum
polarization originates in a circumstellar dust cloud. The observed
wavelength dependence (a power law with polarization significantly
increasing towards the blue) was interpreted as being produced by
small dust grains. In the case of SN~2008S, Botticella et
al. (\cite{mt09}) found evidence of an extended dust shell, which is
understood to be responsible for the large mid-infrared excess
observed in the early phases. This was confirmed by radiative transfer
calculations applied to mid-IR spectra, showing that a substantial
amount of dust must have survived the outburst (Wesson et
al. \cite{wesson}). The existence of a conspicuous dust shell around
NGC~300~OT2008-1, pre-dating the outburst, was demonstrated by Prieto
et al. (\cite{prieto09}) based on a mid-IR Spitzer spectrum obtained
three months after the discovery. All of these results infer that dust
scattering is a perfectly plausible source of continuum polarization
in NGC~300~OT2008-1.

\begin{figure}
\centerline{
\includegraphics[width=80mm]{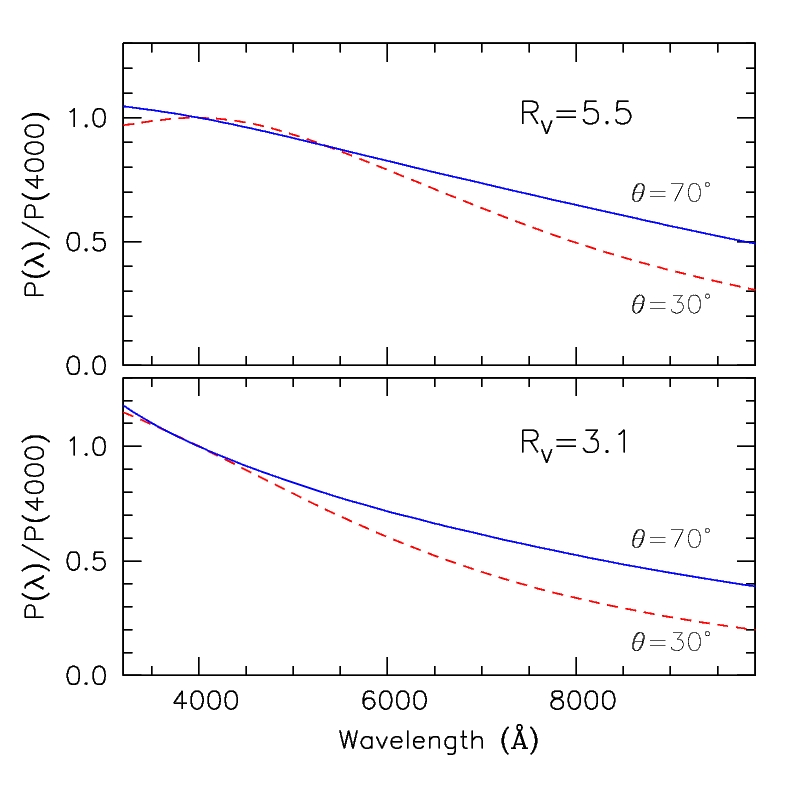}
}
\caption{\label{fig:ds}Expected polarization wavelength dependence of
  a Milky Way carbonaceous-silicate dust mixture in a
  single-scattering approximation for $R_V$=5.5 (upper panel) and
  $R_V$=3.1 (lower panel) and two different scattering
  angles. Dust model is from Li \& Draine (\cite{li}) and Draine
  (\cite{draine03b}).}
\end{figure}

One important problem with the application of the dust-scattering
scenario is the pronounced wavelength dependence this is expected to
produce. This can be easily understood as follows. If there is dust
surrounding the object, part of the light emitted outside the line of
sight would be scattered into the observer's direction and add to the
radiation traveling directly to her. Since the scattering by dust
produces a high degree of linear polarization (see for instance White
\cite{white}), depending on the dust geometrical distribution this can
turn into a significant non null net polarization. If we consider an
infinitesimal dust volume element placed at some distance from the
central source and an incoming unpolarized packet of photons at a
given wavelength, to a first approximation\footnote{If the dust
  optical depth is high, then multiple scattering becomes important
  and acts as a depolarizer. See for instance Patat (\cite{patat05}).}
the fraction of photons scattered into the line of sight (polarized
perpendicularly to the plane of scattering) is proportional to
$C_{ext}(\lambda) \; \omega(\lambda) \; \Phi(\lambda, \theta)$ (White
\cite{white}), where $C_{ext}(\lambda)$ is the dust extinction
cross-section, $\omega(\lambda)$ is the dust albedo, and
$\Phi(\lambda,\theta)$ is the scattering phase function.  This in turn
depends on the scattering angle $\theta$ and the degree of forward
scattering $g(\lambda)$. Since the albedo is roughly constant in the
optical domain and both $C_{ext}$ and $\Phi$ tend to increase
significantly in the blue, this implies that the polarization is
higher at shorter wavelengths. Using the carbonaceous-silicate grain
model by Li \& Draine (\cite{li}) and Draine (\cite{draine03b}) and
adopting the usual parametrization by Henyey \& Greenstein (\cite{hg})
for the scattering phase function, one can estimate this effect more
quantitatively. The result is shown in Fig.~\ref{fig:ds} for two
different scattering angles (30 and 70 degrees) and two different
Milky Way dust mixtures\footnote{Wesson et al. (\cite{wesson}) have
  shown that the dust around SN~2008S most likely consists of carbon
  grains. The usage of a carbonaceous-silicate mixture here is just
  illustrative.}  ($R_V$=3.1, 5.5). As anticipated, the polarization
significantly increases at shorter wavelengths, especially in the case
of lower $R_V$, and can be fairly well reproduced by an exponential
law $P(\lambda)=P_0 \exp[\beta(\lambda-\lambda_0)/\lambda_0]$. For
$\lambda_0$=4000\AA, $\beta$=$-$0.7 for an average scattering angle of
$\theta$=50 degrees (corresponding to the typical value of the forward
scattering degree in the optical domain, i.e., $g\simeq$0.6).

\begin{figure}
\centerline{
\includegraphics[width=80mm]{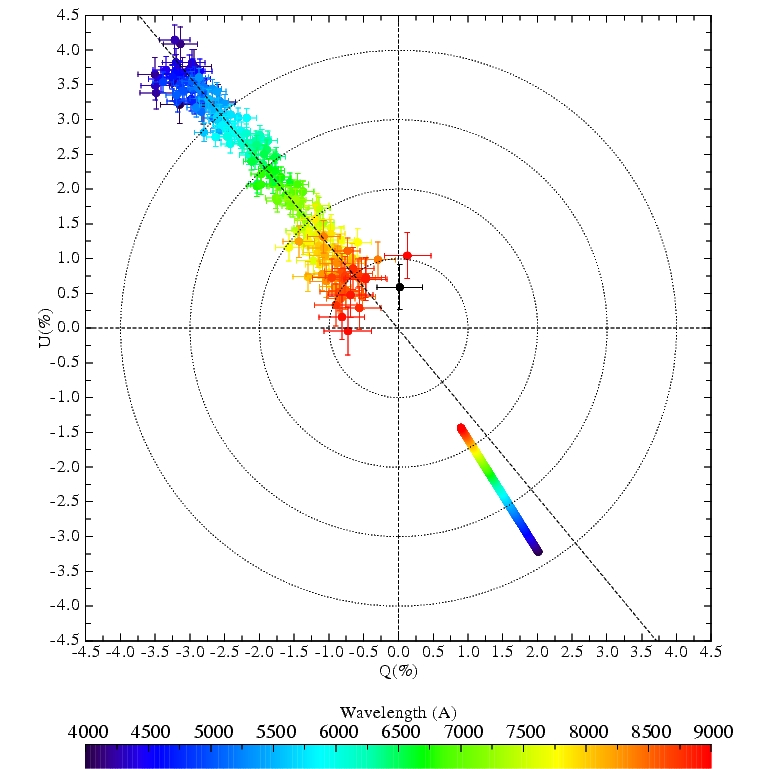}
}
\caption{\label{fig:dustcomp}Minimum polarization solution with an
exponential polarization $\vec{P_1}$ for 2008 July 8. The color-coded
segment in the lower right panel is the exponential component
$\vec{P_1}$, while the dots trace the corresponding component
$\vec{P_2}$. The dashed line traces the best-fit dominant axis.}
\end{figure}

With this result in hand, one can reverse the procedure and, using the
observed data, deduce the implied component $\vec{P_2}$. This is shown
in Fig.~\ref{fig:dustcomp}, where we have vectorially subtracted from
the data the exponential component $\vec{P_1}$ ($P_0$=3.7\%,
$\beta$=$-$0.7 and $\theta_1$=151 degrees). As in the
wavelength-independent case, we have considered the minimum
polarization solution (see Sect.~\ref{sec:cont}). As illustrated by
Fig.~\ref{fig:dustcomp}, the resulting $\vec{P_2}$ is still well
aligned with a dominant axis ($r_{xy}$=$-$0.96): the position angle is
64.7$\pm$0.2 degrees and the RMS deviation is 0.31\%, i.e., slightly
higher than for the wavelength-independent solution
(cf. Table~\ref{tab:dominant}). This is fully consistent with an
alignment along the local spiral arm of NGC~300, which is indicative
of the interstellar nature of this component.

\begin{figure}
\centerline{
\includegraphics[width=80mm]{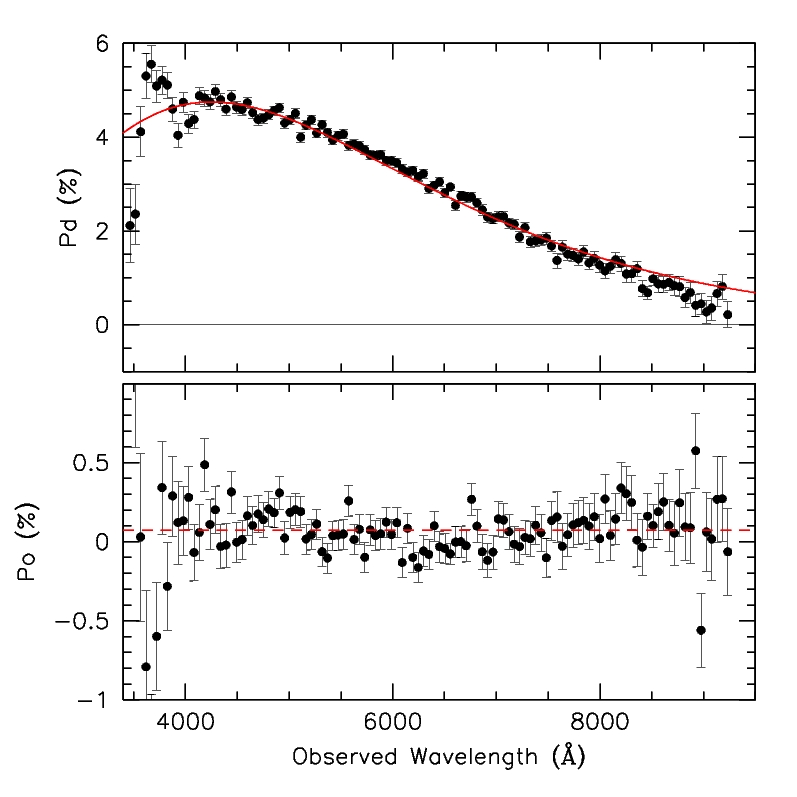}
}
\caption{\label{fig:exprot} Dominant (upper panel) and orthogonal 
(lower panel) polarization on 2008 July 7 after subtracting the
exponential component $\vec{P_1}$.  For presentation, the data were
binned to 16 pixel (57\AA). The solid curve on the upper panel is a
Serkowski law with $P_{max}$=2.6\%, $\lambda_{max}$=4880\AA\/ and
$K$=5.6.}
\end{figure}

The position angle $\theta_1$ was fixed by constraining the dominant
axis to pass through the origin of the $Q-U$ plane (Wang et
al. \cite{wang03}). As done in the wavelength-independent case
(Sect.~\ref{sec:cont}), we decomposed $\vec{P_2}$ along the directions
parallel and perpendicular to the dominant axis, as shown in
Fig.~\ref{fig:exprot} for the second epoch. The RMS deviation of the
orthogonal component $P_o$ is 0.15\%, which is consistent with the
typical measurement errors and indicates that there are no
statistically significant wavelength dependencies in the residual
polarization. As for $P_d$, this shows a smooth behavior, and can be
reasonably well fitted by a Serkowski law with $P_{max}$=4.8\%,
$\lambda_{max}$=4250\AA, and $K$=3.0. This value of $K$ is still
large compared to what is typical of the Milky Way, but it is
significantly smaller than that required by the wavelength-independent
component (Sect.~\ref{sec:cont}).

\begin{figure}
\centerline{
\includegraphics[width=80mm]{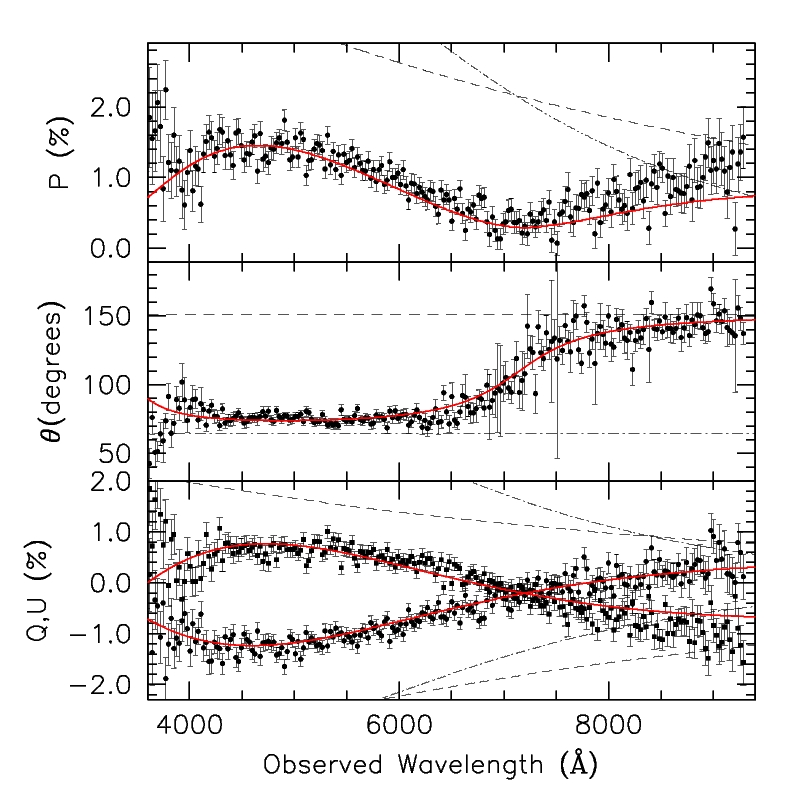}
}
\caption{\label{fig:pexpsol}Best-fit solution for 2008 July 8 
compared to the observed Stokes parameters (lower panel), position
angle (upper panel), and polarization level (upper panel).  The curves
indicate the exponential component $\vec{P_1}$ (dashed),
$\vec{P_2}$ (dotted-dashed), and the global solution (solid).}
\end{figure}

The global solution is compared to the data in Fig.~\ref{fig:pexpsol}
for the second epoch. The fit is reasonably good, even though $\chi^2$
(2.5) and the RMS deviation (0.3\%) are greater than in the
wavelength-independent solution (Sect.~\ref{sec:cont}).  Of course,
there is absolutely no reason why the wavelength dependency of dust
scattering in the circumstellar environment of NGC~300~OT2008-1 should
be similar to that of the Milky Way. However, this illustrates that
the wavelength-independent solution, although giving the most accurate
reproduction of the data with the minimum number of free parameters,
is not unique. One possible argument against a strong wavelength
dependence of $\vec{P_1}$ is that this would imply higher polarization
values in both components. In the example that we have just discussed,
$P_2$ reaches a maximum polarization of about 5\% and $P_1$ is
$\sim$4\% at 4000\AA. These values are quite high, especially
considering that the solutions are undefined up to an additive
constant and so the derived polarization levels are only lower limits.

A detailed treatment of the radiation transfer, including light-travel
effects, multiple scattering, dust properties, and geometrical
distribution (see Patat \cite{patat05}) would be required for a
realistic modeling. However, a lower limit to the amount of dust
required to account for the observed level of polarization can be
obtained as follows. We assume that the dust is placed at a distance
$r$ from the supposedly unpolarized source and is distributed in such
a way that all scattered radiation becomes polarized to the maximum
possible level $P_{max}$ (attained for $\theta$=$\pi$/2 and
$P_{max}\approx$0.5; see White \cite{white}). Based on this
assumption, if $F_s(\lambda)$ is the flux scattered into the line of
sight and $F_0(\lambda)$ is the flux isotropically emitted by the
source, then the observed degree of polarization is given by

\begin{displaymath}
P(\lambda)\approx \frac{F_s}{F_0+F_s} \approx \frac{P_{max}}{4\pi r^2} \; 
\Phi(\theta,\lambda) \; \omega(\lambda) \; C_{ext}(\lambda) \; 
\frac{M_H}{m_H},
\end{displaymath}

\noindent where $M_H$ is the implied hydrogen mass (for a given
gas-to-dust mass ratio) and $m_H$ is the proton mass. Substituting the
values for the $R_V$=3.1 dust mixture (Li \& Draine \cite{li}; Draine
\cite{draine03b}) for $\lambda$=5500\AA\/ ($\omega$=0.67, $g$=0.54,
$C_{ext}/H$=5$\times$10$^{-22}$ cm$^2$) and assuming maximum
polarization ($\theta$=$\pi$/2), one arrives at the
expression:

\begin{displaymath}
\frac{M_H}{M_\odot}\geq 3.8\times10^{-7} \left( \frac{r}{AU} \right)^2\;P(5500).
\end{displaymath}

\noindent For instance, if the scattering material is placed at
$r$=1000 AU, one needs $M_H\approx$8$\times$10$^{-3}$ $M_\odot$ to
reach a 2\% polarization. For a gas-to-dust mass ratio of 200 (typical
of evolved stars with carbon-rich dust; see Matsuura et
al. \cite{matsuura}), this implies a lower limit to the dust mass of
4$\times$10$^{-5}$ $M_\odot$. Clearly, if the dust is placed at 100
AU\footnote{Prieto et al. (\cite{prieto09}) set the outer boundary of
  dust evaporation for NGC~300~OT2008-1 to be 100 AU, while for
  SN~2008S Botticella et al. (\cite{mt09}) obtained a significantly
  larger value ($\sim$2000 AU).}, this lower limit would decrease by a
factor of 100.  As we said, this is only the mass of dust distributed
in an asymmetric way, because polarization alone cannot constrain the
total amount of dust. For instance, a spherically symmetric dust shell
of 10$^{-2}$ $M_\odot$ could be present, but this would not produce
any measurable polarization. Prieto et al. (\cite{prieto09}) came to
the conclusion that about 10$^{-4}$ $M_\odot$ of circumstellar dust
are required to explain the mid-IR excess in NGC~300~OT2008-1. Similar
results were found by Wesson et al. (\cite{wesson}) for SN~2008S,
while Botticella et al. (\cite{mt09}) derived a dust mass of
$\sim$10$^{-3}$ $M_\odot$.  Our data suggest that a significant
fraction of this material (the exact amount depends on the dust
distance among other parameters) must be distributed in an asymmetric
configuration. Whether this is a torus, a disk, a bipolar outflow (see
Bond et al. \cite{bond09}; Prieto et al. \cite{prieto09}) or a clumpy
shell superimposed on an otherwise spherically symmetric distribution,
we are unable to decide on the basis of the available
data. Spectropolarimetry obtained during earlier epochs might help to
differentiate between the various possibilities.

The fiducial date of the outburst is assumed to be April 17, 2008
(Berger et al. \cite{berger09}), i.e., 75 days before our first
epoch. This implies that, at the time of our observations, the
expanding radiation front is illuminating dust placed at $\sim$10$^4$
AU perpendicular to the line of sight. The light curve peak has a
width of about 100 days (Bond et al. \cite{bond09}), so that the
polarized photons that we have detected were scattered in a region
that extends between the dust evaporation boundary and about 10$^4$
AU. Given that at this large distance the incoming flux is a factor of
100 lower than at 1000 AU, the bulk of the polarization probably comes
from material located closer to the source, between the dust
evaporation radius and a few thousand AU. In this scenario, the
observed decrease in polarization can be understood as being caused by
the expansion of the radiation flash, coupled to the confinement of
dust close to the object and possible inhomogeneities in its
distribution.

If the dust is distributed across a sufficiently large distance from
the outbursting object perpendicular to the line of sight, so that the
scattering material produces a long integration time, then a case B
light echo (Patat et al. \cite{patat06b}) might become observable as
the transient fades out.  Its evolution with time could help us to
better characterize the dust distribution.  Unfortunately, the lack of
a scattered light echo detection in the HST images obtained more than
4 months after the outburst (Bond et al. \cite{bond09}) does not allow
us to place stringent limits on the circumstellar dust distance. The
HST-ACS spatial resolution (50mas) corresponds to $\sim$10$^5$ AU at
the distance of NGC~300 (1.88 Mpc; Gieren et al. \cite{gieren}; Rizzi
et al. \cite{rizzi}). Future HST observations might clarify whether
there is circumstellar dust at distances greater than $\sim$10$^5$ AU
from the outburst site. Incidentally, the absence of a resolved light
echo indicates that the transient must be placed at a significant
distance behind the interstellar dust associated with the local spiral
arm of NGC~300.

\section{\label{sec:conc}Conclusions}

We have presented VLT-spectropolarimetry of NGC~300~OT2008-1 obtained
48 and 55 days after its discovery.  The continuum polarization is
dominated by a component very well aligned along the local spiral arm
of NGC~300, which we have interpreted as originating in interstellar
dust. The wavelength dependency of the ISP differs substantially from
what is typical of the Milky Way.  Once the ISP is subtracted, a
substantial signal, reaching about 1.2\%, is detected. This
polarization component is aligned along a completely different position
angle.

The simplest explanation of this continuum polarization is electron
scattering coupled with a significantly asymmetric
photosphere. Nevertheless, the absence of any line depolarization
(very common in sources displaying similar levels of polarization)
disfavors this scenario and calls for an alternative origin.

Although we have investigated different possible reasons for the lack
of depolarization, the most viable physical mechanism is scattering by
a circumstellar, asymmetric dust cloud. If this is the correct picture
(also implying that the emitting photosphere does not deviate
significantly from spherical symmetry), this means that a significant
amount of dust ($\geq$10$^{-5}$ $M_\odot$) must be present during the
epochs covered by our observations, either because it has re-condensed
or because it has survived the radiation flash produced by the
outburst.

Besides adding to the growing evidence of substantial amounts of
circumstellar dust in a completely independent way, our observations
provide direct proof that dust is not only present, but that a
significant fraction must have an asymmetric geometry. This implies
that one or more asymmetric outflow episodes from the progenitor star
must have occurred during its past history.

\appendix

\section{\label{sec:app}Effects of second order contamination on 
spectropolarimetry}

The purpose of this Appendix is to evaluate the effect of the second
order on spectropolarimetry when no order-sorting filter (hereafter
OSF) is used, and to propose a simple technique for its correction.

\subsection{\label{sec:app1}Second-order contamination}

In general, diffraction-based dispersive elements (such as gratings and
grisms) produce replicas of the first-order (FO) spectrum with
dispersions that increase proportionally to the order number. While
higher orders can usually be neglected, the second-order (SO) spectrum
appears as an additive component superimposed on the red part of the
first order (FO) spectrum, with a dispersion that is $\sim$2 times
that of the first order. In practice, the flux detected at a given
wavelength $\lambda$ is the sum of the {\it clean} spectrum at
$\lambda$ and a fraction of the spectrum detected at a bluer
wavelength, which we indicate as $\lambda^\prime$. If
$s(\lambda)$ is the spectrum one would record on the detector in the
absence of SO contamination, the spectrum which is actually observed
$s^\prime(\lambda)$ can be modeled as

\begin{equation}
\label{eq:soc}
s^\prime(\lambda) = s(\lambda) + r_{2/1}(\lambda) \; s(\lambda^\prime)
\end{equation}

\noindent where $r_{2/1}(\lambda)$ is a function that we wish to
derive.  The first step in this procedure is to derive the relationship
between $\lambda$ and $\lambda^\prime$. For this purpose, we
used archival FORS1 wavelength-calibration exposures taken with
and without the OSF GG435 (O'Brien \cite{fors}). After performing the
wavelength calibration and identifying 6 isolated blue lines and their
red replicas, we derived the relation between their first order
wavelength and their apparent wavelength in the second order. The data
points are very well fitted by a linear relation

\begin{equation}
\label{eq:lambda}
\lambda = a + b\;\lambda^\prime,
\end{equation}

\noindent where $a$=$-$539.5$\pm$0.4\AA\/ and
$b$=2.081$\pm$0.001. With this result at hand, the ratio $r_{2/1}$ can
be estimated in the following way. We indicate by $S(\lambda)$ the
intrinsic spectrum of an astronomical source, by $T(\lambda)$ the
transmission function (including atmosphere, instrument optics and
detector efficiency), by $F(\lambda)$ the OSF transmission, and by
$G_1(\lambda)$ and $G_2(\lambda)$ the grism FO and SO order
efficiencies,respectively.  If $s_{f}(\lambda)$ and $s_{nf}(\lambda)$
are the observed spectra obtained with and without OSF, respectively,
and based on the assumption that the order-sorting filter completely
eliminates the SO (see below), they can be written as

\begin{displaymath}
\begin{array}{lll}
s_{f}(\lambda) & = & S(\lambda) \;T(\lambda) \; F(\lambda) \;G_1(\lambda)\\
s_{nf}(\lambda)& = & S(\lambda) \;T(\lambda) \;G_1(\lambda) + 
S(\lambda^\prime) \;T(\lambda^\prime) \;G_2(\lambda^\prime).\\
\end{array}
\end{displaymath}

Since $r_{2/1}$, as defined in Eq.~\ref{eq:soc}, is the ratio of
the {\it pure} SO flux at $\lambda$ to the corresponding FO flux at
$\lambda^\prime$, this can be written as

\begin{equation}
\label{eq:ratio}
r_{2/1} = \frac{s_{nf}(\lambda)-s_f(\lambda)/F(\lambda)}{s_{nf}(\lambda^\prime)}.
\end{equation}

\begin{figure}
\centerline{
\includegraphics[width=90mm]{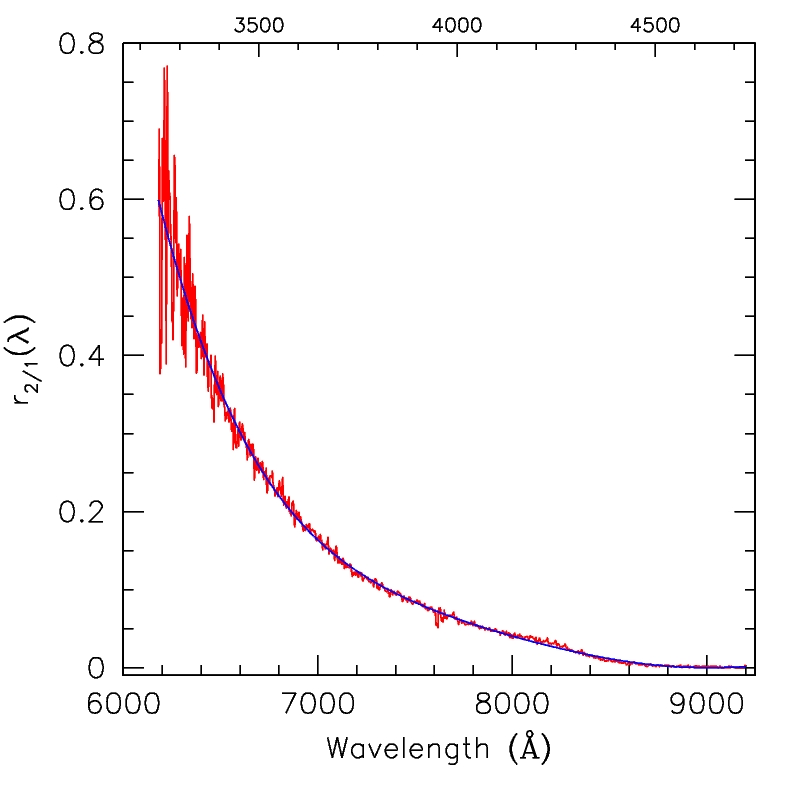}
}
\caption{\label{fig:soc}Ratio between observed SO and FO as a function of 
wavelength derived for the 300V grism mounted in FORS1. The upper
scale indicates the corresponding FO wavelength $\lambda^\prime$. The
solid blue curve is a best fit with a fifth order polynomial.}
\end{figure}

\noindent The advantages of this formulation are that it is based on
measurable or known quantities and is independent of the detector
sensitivity function. It can indeed be easily shown that
$r_{2/1}(\lambda)$ is the ratio of $G_2$ to $G_1$.

In principle, any bright astronomical target with a continuum spectrum
observed with and without the OSF can be used to derive
$r_{2/1}(\lambda)$. Nevertheless, because of the lower sensitivity of
detectors in the blue and since the effect that we wish to measure is
produced by blue photons, the most suitable choice is a hot star.  For
this reason, and because it is included in the FORS1 calibration plan
(O'Brien \cite{fors}), we selected the D0p spectrophotometric standard
star Feige~110 ($\alpha_{2000}$ = 23:19:58.4,
$\delta_{2000}$=$-$05:09:55.8), which has a very blue ($B-V$=$-$0.30)
and featureless spectrum (Hamuy et al. \cite{hamuy92},
\cite{hamuy94}).  The FORS1 archival data that we selected were
obtained on 2008 October 12 with the grism 300V, with and without the
OSF GG435, which has a cutoff wavelength $\lambda_f\sim$4300\AA. As
for the filter transmission curve, we used the tabulated data provided
by the
Observatory\footnote{http://www.eso.org/sci/facilities/paranal/instruments/fors/inst/Filters/}. One
needs to mention that the GG435 OSF does not prevent a small portion
of the SO being produced in the red edge of the spectrum. Its blue
cut-off corresponds to a SO wavelength of $\sim$8500\AA, so that
redward of this value a contribution from photons bluer than
4350\AA\/ is still present. This makes the derivation of
$r_{2/1}(\lambda)$ uncertain between 8500\AA\/ and the red edge of the
spectral range covered by the detector.

After wavelength calibration, we extracted the two spectra and
applied Eq.~\ref{eq:ratio} to derive $r_{2/1}(\lambda)$. The
result is presented in Fig.~\ref{fig:soc}. This figure shows that the
second-order contamination can be significant for very blue targets:
for instance, the SO contamination at 6500\AA\/ is about 35\% of the
flux detected at 3380\AA. Clearly, the exact amount of contamination
depends on the input spectrum, which is in principle
unknown. Nevertheless, one must notice that for the SO contamination
only the knowledge of the blue part of the spectrum is needed, where
the observed and the uncontaminated spectrum are identical
($s^\prime(\lambda)=s(\lambda)$). The SO begins to contaminate
the spectrum at a wavelength $\lambda_c$ which corresponds to the
detector blue cut-off wavelength $\lambda_b$. For FORS1, it is
$\lambda_b\sim$3200\AA, which, according to Eq.~\ref{eq:lambda},
corresponds to $\lambda_c$=6120\AA. Below this wavelength, the SO can
be considered negligible and so one can set $r_{2/1}(\lambda)$=0 for
$\lambda<\lambda_c$.

Therefore, knowing $r_{2/1}(\lambda)$ one can recover the uncontaminated
spectrum by simply inverting Eq.~\ref{eq:soc} and using
Eq.~\ref{eq:lambda}, e.g.,

\begin{equation}
\label{eq:corr}
s(\lambda)=s^\prime(\lambda) - r_{2/1}(\lambda) \; 
s^\prime\left(\frac{\lambda-a}{b}\right).
\end{equation}

\noindent A way of expressing the severity of the SO contamination is to
normalize its absolute amount to the flux at any given wavelength,
by introducing the quantity $\Delta(\lambda)$

\begin{equation}
\label{eq:delta}
\Delta(\lambda) = \frac{s^\prime(\lambda)-s(\lambda)}{s^\prime(\lambda)} =
r_{2/1}(\lambda)\; \frac{s^\prime(\lambda^\prime)}{s^\prime(\lambda)},
\end{equation}

\noindent which is the fraction of the observed flux produced by the SO. 

\begin{figure}
\centerline{
\includegraphics[width=90mm]{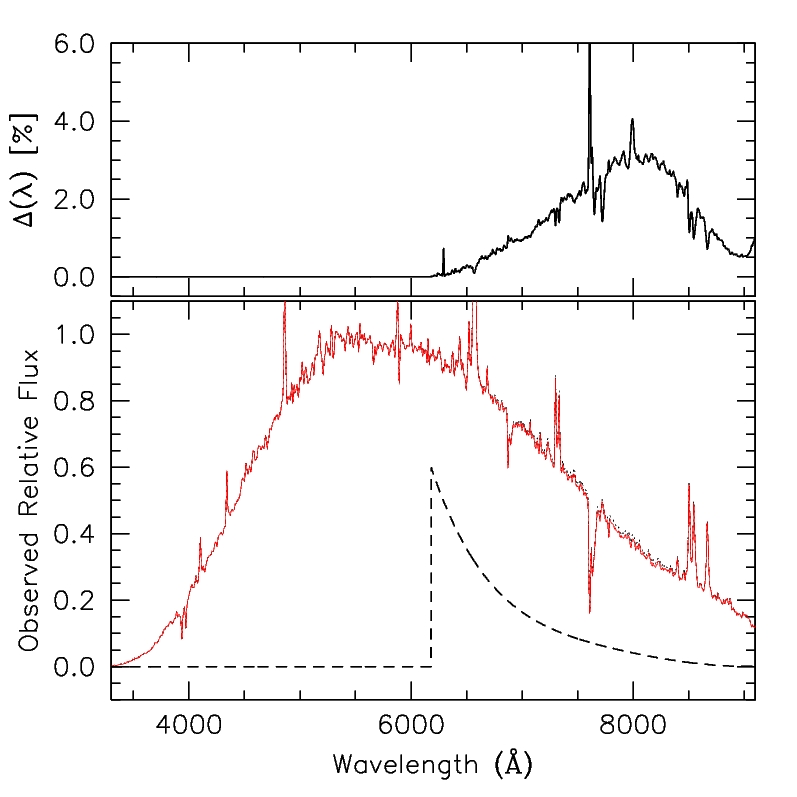}
}
\caption{\label{fig:contam}Lower panel: comparison between the observed 
spectrum of the OT in NGC~300 (dotted line) and the corrected spectrum
(solid line). The dashed line traces the $r_{2/1}(\lambda)$ function.
Upper panel: the relative SO contamination $\Delta(\lambda)$. The
results above 8500\AA\/ are uncertain (see text).}
\end{figure}

The application of this method to the data discussed in this paper is
illustrated in Fig.~\ref{fig:contam}. To avoid introducing additional
noise, we replaced the computed $r_{2/1}(\lambda)$ with a fifth
order polynomial fit to the data (solid curve in
Fig.~\ref{fig:soc}). For our object, the contamination reaches a
maximum relative value of about $\Delta$=5\% at about 8000\AA\/ which,
for most purposes is negligible, being smaller than the typical
accuracy one can achieve in the flux calibration.

Nevertheless, since polarimetry is in most cases dealing with flux
differences of a few percent, the impact of SO contamination needs
to be evaluated.

\subsection{\label{sec:app2}Effects on spectropolarimetry}

In dual-beam instruments such as FORS1, spectropolarimetry is performed
by evaluating the so-called normalized flux ratios $F_i(\lambda)$, derived
from the ordinary and extraordinary beams produced in the same frame
by the Wollaston prism (see Patat \& Romaniello \cite{patat06a} for a
general introduction). For the canonical HWP position angles
$\theta_i=\frac{\pi}{8} i$ (where $i$=0, 1, 2, 3) and for an ideal
polarimeter, these ratios are directly related to the Stokes
parameters: $F_0=Q$, $F_1=U$, $F_2=-Q$, and $F_3=-U$. Since the
observed fluxes in the ordinary ($f^{\prime,O}_i$) and extraordinary
($f^{\prime,E}_i$) beams can be modeled with Eq.~\ref{eq:soc}, the
observed normalized ratios

\begin{displaymath}
F^\prime_i(\lambda)=\frac{f^{\prime,O}_i(\lambda)-f^{\prime,E}_i(\lambda)}
{f^{\prime,O}_i(\lambda)+f^{\prime,E}_i(\lambda)}
\end{displaymath}

\noindent after some manipulation can be expressed as

\begin{equation}
\label{eq:F}
F^\prime_i(\lambda)=\frac{F_i(\lambda) + g_i(\lambda)\; F_i(\lambda^\prime)}
{1+g_i(\lambda)},
\end{equation}

\noindent where we have defined

\begin{displaymath}
g_i(\lambda)=\frac{f^O_i(\lambda^\prime)+f^E_i(\lambda^\prime)}
{f^O_i(\lambda)+f^E_i(\lambda)} \; r_{2/1}(\lambda)
\equiv \frac{s_i(\lambda^\prime)}{s_i(\lambda)}\; r_{2/1}(\lambda)
\end{displaymath}

Equation~\ref{eq:F} shows that, in general, the effect of SO
contamination on the final polarization is not obvious, since it
depends on the exact spectral energy distribution and polarization of
the source.  However, when the SO contamination is moderate
($\Delta(\lambda)\leq$0.1), the effect is more readily understood. In
those circumstances, it is given by $s(\lambda^\prime)/s(\lambda)
\approx s^\prime(\lambda^\prime)/s^\prime(\lambda)$, so that
$g(\lambda)\approx \Delta(\lambda)$ (see Eq.~\ref{eq:delta}) and one
can write

\begin{displaymath}
F^\prime_i(\lambda)\approx\frac{F_i(\lambda) + \Delta(\lambda)\; 
F_i(\lambda^\prime)}{1+\Delta(\lambda)}
\approx F_i(\lambda) + \Delta(\lambda)\;F_i(\lambda^\prime).
\end{displaymath}

\noindent In our case, where $\Delta$ reaches a maximum value of
$\sim$0.05 and the polarization in the blue is of the order of 2\%,
the absolute effect on the derived Stokes parameters is of the order
of 0.1\%, which is comparable to the maximum accuracy one can reach
with instruments such as FORS1 (Patat \& Romaniello \cite{patat06a}). For
this reason, the SO effect on the data presented in this paper can be
neglected.  A more rigorous calculation performed by correcting each of
the ordinary and extraordinary beams following the procedure outlined
in Sect.~\ref{sec:app1} leads to the same conclusions.

\section{\label{sec:appb}The effect of multiple weak polarizers}

In general, if $\vec{S_0}(I_0,Q_0,U_0,V_0)$ is the Stokes vector that
defines the polarization status of an incoming beam, its passage
through an optical system can be described by $\vec{S_1}=\vec{M} \cdot
\vec{S_0}$, where $\vec{S_1}$ is the emerging Stokes vector and
$\vec{M}$ is the Mueller matrix that characterizes the system
(Chandrasekhar \cite{chandra}). To a first approximation, a dust cloud
can be treated as a partial polarizer. The Mueller matrix for a
non-birefringent, non-scattering, partial polarizer with horizontal
transmission axis ($\theta$=0) is given by the expression

\begin{equation}
\label{eq:muel1}
\vec{M(0)}=\frac{1}{2}
\left( \begin{array}{cccc}
k_1+k_2 & k_1-k_2 & 0               & 0 \\
k_1-k_2 & k_1+k_2 & 0               & 0 \\
0       & 0       & 2\sqrt{k_1 k_2} & 0 \\
0       & 0       & 0               & 2\sqrt{k_1 k_2} \\ 
\end{array} \right)
\end{equation}

\noindent where $k_1$ and $k_2$ ($k_1\geq k_2$; 0$\leq k_1,k_2 \leq$1) are the
principal transmittances (see for instance Keller \cite{keller},
Eq.~3.24). For $k_1$=1 and $k_2$=0, one obtains a perfect linear,
horizontal polarizer, while for $k_1=k_2$=1, Eq.~\ref{eq:muel1} implies
an isotropic, non-absorbing medium. The general expression for a
generic position angle $\theta$ can be obtained by computing the product
$\vec{R(-\theta)} \cdot \vec{M(0)}
\cdot
\vec{R(\theta)}$ (Chandrasekhar \cite{chandra}),  where $\vec{R(\theta)}$ is the
usual rotation matrix

\begin{displaymath}
\label{eq:rot}
\vec{R(\theta)}=
\left( \begin{array}{cccc}
1 & 0           & 0          & 0 \\
0 & \cos 2\theta  & \sin 2\theta & 0 \\
0 & -\sin 2\theta & \cos 2\theta & 0 \\
0 & 0           & 0          & 1\\ 
\end{array} \right).
\end{displaymath}

\noindent After defining $s=\sin 2\theta$, $c=\cos 2\theta$,
$k=(k_1+k_2)/2$, and $p$=$(k_1-k_2)/(k_1+k_2)$, one can write the
matrix product as

\begin{displaymath}
\label{eq:muel2}
\vec{M(\theta)}= k
\left( \begin{array}{cccc}
1  & pc                   & ps                  & 0 \\
pc & c^2+s^2\sqrt{1-p^2}  & (1-\sqrt{1-p^2})sc  & 0 \\
ps & (1-\sqrt{1-p^2})sc   & s^2+c^2\sqrt{1-p^2} & 0 \\
0  & 0                    & 0                   & \sqrt{1-p^2}\\ 
\end{array} \right),
\end{displaymath}

\noindent where $0\leq k,p \leq1$. For $p$=0, one clearly obtains an
isotropic medium with transmittance $k$, while for $p$=1 the previous
equation provides the general expression of an ideal linear polarizer
with position angle $\theta$ (see for instance Keller \cite{keller};
Eq.~3.26).

Using the above matrix, one can compute the resulting Stokes vector to
be

\begin{displaymath}
\label{eq:trans}
\begin{array}{ccl}
I_1/k &=& I_0+Q_0pc + U_0ps,\\
Q_1/k &=& I_0pc + Q_0\left(c^2+s^2\sqrt{1-p^2}\right) 
		+ U_0\left(1-\sqrt{1-p^2}\right)sc,\\
U_1/k &=&  I_0ps+ Q_0\left(1-\sqrt{1-p^2}\right)sc
                + U_0\left(s^2+c^2\sqrt{1-p^2}\right), \\
V_1/k &=& V_0 \sqrt{1-p^2}.\\
\end{array}
\end{displaymath}

\noindent If the incoming beam is unpolarized, the resulting Stokes
vector is $\vec{S_1}(kI_0, kI_0 p \cos 2\theta, kI_0 p \sin 2\theta,
0)$, so that the polarization is $p$ and its position angle is
$\theta$. If the beam is polarized (either because the source is
intrinsically polarized or because the beam passed already through a
polarizing system such as a dust cloud), then the situation becomes
more complicated.  However, when the incoming polarization and $p$ are
far smaller than 1, the previous expressions can be approximated to be

\begin{displaymath}
\label{eq:approx}
\begin{array}{ccl}
I_1 &\approx& k I_0\\
Q_1 &\approx& k Q_0 + k I_0p\cos 2\theta\\
U_1 &\approx& k U_0 + k I_0p\sin 2\theta\\
V_1 &\approx& k V_0.\\
\end{array}
\end{displaymath}

\noindent Therefore, in the case of weak linear polarization, the output
normalized Stokes parameters are simply given by

\begin{displaymath}
\label{eq:norm}
\begin{array}{ccl}
<Q_1> &\approx& <Q_0> +  p\cos 2\theta,\\
<U_1> &\approx& <U_0> +  p\sin 2\theta.\\
\end{array}
\end{displaymath}

\noindent This implies that the effect of multiple weak polarizers (such as
several dust clouds) can be described as a sum of terms, each
characterized by a polarization $p_i$ and a position angle
$\theta_i$. In other words, when the polarization is weak, the
components can be considered to be vectorially additive\footnote{The
  demonstration can be easily extended to the most general form of the
  Mueller matrix that describes the polarizer.}.

An interesting consequence follows from this conclusion. If we express
the polarization wavelength dependency as $p(\lambda)=p_{max}
f(\lambda)$ and we imagine a beam of unpolarized light passing through
a system of interstellar clouds of various $p_{max}$ and $\theta$
(with $d\theta/d\lambda$=0), then the resulting normalized Stokes
parameters are

\begin{displaymath}
\begin{array}{lcl}
<Q> &=& \sum p_{max,i} \; f_i(\lambda) \;\cos 2\theta_i, \\
<U> &=& \sum p_{max,i} \; f_i(\lambda) \;\sin 2\theta_i. \\
\end{array}
\end{displaymath}

\noindent Therefore, if the dust in the various clouds produces linear
polarization with the same wavelength dependency (i.e., dust properties
are common to all clouds), then the resulting polarization position
angle $\theta=1/2 \arctan (U/Q)$ is wavelength-independent and the entire
system is indistinguishable from a single cloud.

On the other hand, if the different clouds produce polarizations that
do not obey the same law $f(\lambda)$, then the resulting
polarization position angle $\theta$ will be
 wavelength-dependent. Once plotted on the $Q,U$ plane, the
polarization data will not lie on a straight line passing through the
origin and will possibly extend over different quadrants.

\begin{figure}
\centerline{
\includegraphics[width=90mm]{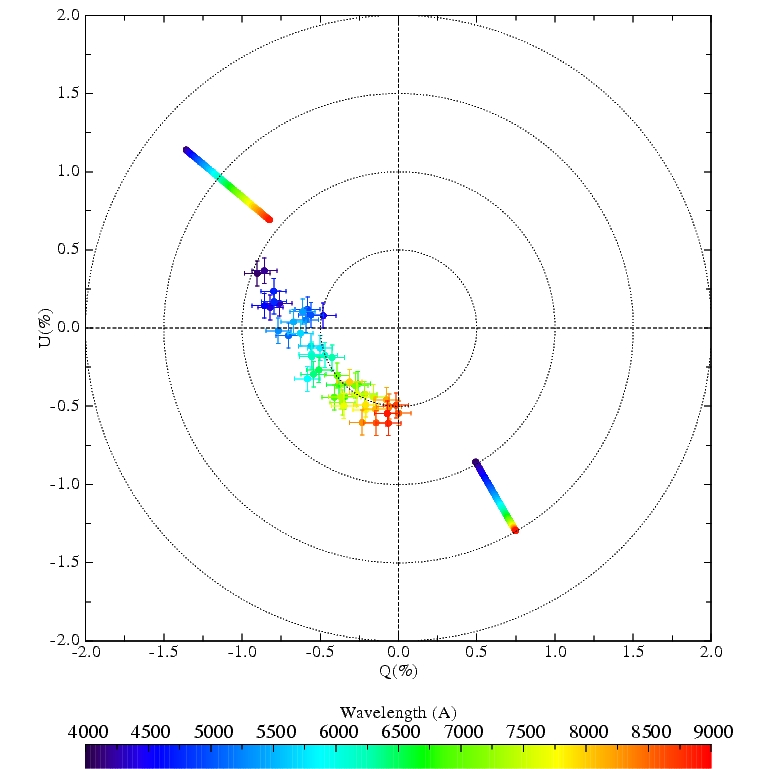}
}
\caption{\label{fig:twocomp}Example composition of two Serkowski laws:
$P_{max}$=1.8\%, $\lambda_{max}$=3500\AA, $\theta$=70$^\circ$ (K=0.59)
and $P_{max}$=1.5\%, $\lambda_{max}$=9500\AA, $\theta$=150$^\circ$
(K=1.59). A Gaussian noise ($\sigma$=0.1\%) was added to the
resulting points. The wavelength step is 100\AA.}
\end{figure}

An example of a system consisting of two clouds both obeying the
empirical Serkowski law (Serkowski et al. \cite{serkowski}), with
two different values of the polarization peak wavelength
$\lambda_{max}$ is shown in Fig.~\ref{fig:twocomp}.

\begin{acknowledgements}
This paper is based on observations made with ESO Telescopes at
Paranal Observatory under program ID 281.D-5016. The authors wish to
thank ESO's Director General for granting his Discretionary Time to
this project. They are also grateful to A. Pastorello and P. H\"oflich
for their kind and competent help, and to H.M. Schmid for the very
useful comments provided during the refereeing process.
\end{acknowledgements}

\end{document}